\documentclass[12pt]{article}

\title{Software for Computing the Spheroidal Wave Functions Using Arbitrary Precision Arithmetic}
\author{Ross Adelman, Nail A. Gumerov, and Ramani Duraiswami}
\date{}

\usepackage{amsmath}
\usepackage{amsfonts}
\usepackage{amsthm}
\usepackage[bf,labelsep=period]{caption}
\usepackage[margin=1.0in]{geometry}
\usepackage{microtype}
\usepackage[pdftex]{graphicx}
\usepackage{verbatim}

\newcommand{\psum}{\sideset{}{^\prime}{\sum}}

\begin{document}

\maketitle

\section*{Abstract}

The spheroidal wave functions, which are the solutions to the Helmholtz equation in spheroidal coordinates, are notoriously difficult to compute.
Because of this, practically no programming language comes equipped with the means to compute them.
This makes problems that require their use hard to tackle.
We have developed computational software for calculating these special functions.
Our software is called \verb#spheroidal# and includes several novel features, such as: using arbitrary precision arithmetic; adaptively choosing the number of expansion coefficients to compute and use; and using the Wronskian to choose from several different methods for computing the spheroidal radial functions to improve their accuracy.
There are two types of spheroidal wave functions: the prolate kind when prolate spheroidal coordinates are used; and the oblate kind when oblate spheroidal coordinate are used.
In this paper, we describe both, methods for computing them, and our software.
We have made our software freely available on our webpage.

\section{Introduction}

Elementary functions, such as powers, sines, and exponentials, are solutions to differential equations arising in various math, science, and engineering problems.
Other elementary functions, such as roots, arcsines, and logarithms, are inverses of these.
They are called elementary because the differential equations from which they are derived are usually linear, homogenous, and have constant coefficients.
This makes them easy to solve and the expressions for their solutions easy to compute.
In many problems, however, the differential equations encountered are not so easy to solve, and their solutions can be extremely complicated.
The functions arising from these are called special functions.
Examples of such functions are the associated Legendre polynomials and the spherical Bessel functions.

In this paper, we explore the spheroidal wave functions.
These special functions are the solutions to the differential equations obtained by applying the method of separation of variables to the Helmholtz equation in spheroidal coordinates.
There are two cases: the prolate spheroidal wave functions when prolate spheroidal coordinates are used; and the oblate spheroidal wave functions when oblate spheroidal coordinates are used.
Unfortunately, there are no simple expressions for computing them.
Instead, they must be written as infinite series expansions in terms of various other special functions.
For example, the spheroidal angle functions can be written in terms of the associated Legendre polynomials, and the spheroidal radial functions can be written in terms of the spherical Bessel and Neumann functions.
Depending on the method used, there are three or four different sets of expansion coefficients that need to be computed.

The spheroidal wave functions have applications in many disciplines.
Our primary motivation for studying them was for computing the solutions to acoustic scattering problems involving prolate spheroids, oblate spheroids, and disks \cite{bowman1987}.
However, they are also encountered in signal processing \cite{slepian1983}.

We have developed computational software for calculating the spheroidal wave functions using C++ and MATLAB.
Our software is called \verb#spheroidal# and includes the following features.
First, the software uses GNU MPFR, a library for performing arbitrary precision arithmetic \cite{fousse2007}.
Using arbitrary precision arithmetic provides greater accuracy in many of the computations, especially for higher wavenumbers and modes.
Second, the software allows the user to specify the level of precision to use at every computational step.
For example, different levels of precision can be used for computing the different sets of expansion coefficients that need to be computed.
Third, the software allows the user to specify in two different ways how many expansion coefficients to compute and use.
In the first way, the user specifies the number of expansion coefficients exactly (e.g., compute 200 of this type and 300 of that type).
In the second way, the user allows the software to choose the number of expansion coefficients adaptively.
All of the expansion coefficients decay exponentially in the long run, so in this method, the user specifies the minimum magnitude that the expansion coefficients should reach (e.g., keep computing expansion coefficients until the next one drops below $10^{-200}$).
Fourth, the expansion coefficients, as well as other special values, are saved to disk.
This way, they can be reused later on without having to recompute them from scratch.
Fifth, there are several methods for computing the spheroidal radial functions.  The actual value of the Wronskian of these functions is easy to compute, so the combination of methods for computing the spheroidal radial functions is chosen so that the computed Wronskian has the smallest error.
We have made our software freely available for download on our webpage.

The spheroidal wave functions have been studied for over six decades.
Perhaps the most complete description of the spheroidal wave functions is given by Flammer \cite{flammer2005}.
As pointed out by \cite{do-nhat1996}, some of the expressions in \cite{flammer2005} are incorrect.
Nevertheless, Flammer's book is an invaluable resource.
Actually implementing the spheroidal wave functions in code is very involved.
In \cite{zhang1996}, they were implemented in Fortran, and in \cite{thompson1997, thompson1999}, they were implemented in C.
These implementations used double precision.
Due to round-off errors, double precision can lead to large errors, especially for higher frequencies and modes.
In \cite{vanburen2002, vanburen2004}, they were implemented in Fortran using quad precision and expressions that converge faster and more accurately in some cases to obtain better accuracy over a wide range of frequencies, modes, and argument values.
These implementations, as well as our own, are only for real frequnecies and integer modes.
Other authors have investigated complex frequnecies and non-integer modes \cite{li1998, falloon2003}, and in these particular references, the respective authors used Mathematica, which can work in arbitrary precision.
Some have also looked at numerical techniques, such as finite difference approximations and relaxation methods \cite{caldwell1988, ogburn2014}.
Many of these authors have released their code free to use.

\section{Spheroidal Coordinates}

\begin{figure}[t]
	\centering
	\includegraphics[height=2.4in]{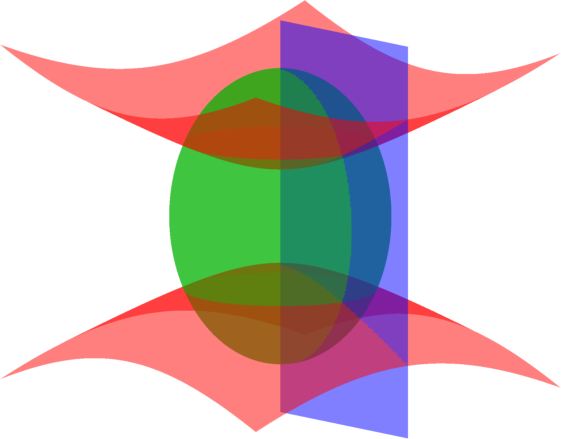}
	\includegraphics[height=2.4in]{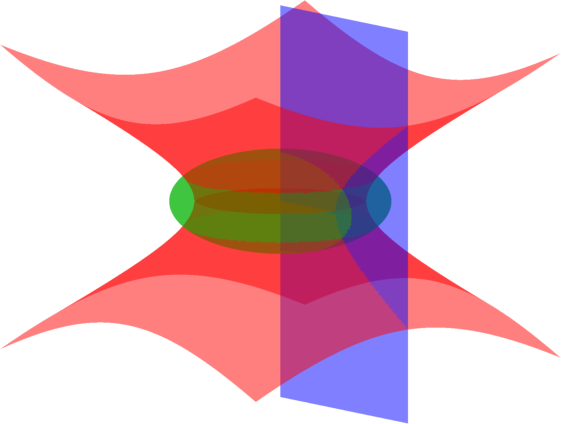}
	\caption{The prolate (left) and oblate (right) spheroidal coordinate systems.  The three colored surfaces are isosurfaces for $\eta = \pm1 / 2$ (red), $\xi = 3 / 2$ for the prolate case and $\xi = 1 / 2$ for the oblate case (green), and $\phi = 0$ (blue).}
	\label{y001}
\end{figure}

The prolate spheroidal coordinate system, shown in Figure \ref{y001}, is related to the Cartesian coordinate system by \cite{flammer2005}
\begin{equation}
x = a\left(1 - \eta^2\right)^{1 / 2}\left(\xi^2 - 1\right)^{1 / 2}\cos\left(\phi\right),\quad
y = a\left(1 - \eta^2\right)^{1 / 2}\left(\xi^2 - 1\right)^{1 / 2}\sin\left(\phi\right),\quad
z = a\eta\xi,
\end{equation}
where $2a$ is the interfocal distance.
The Helmholtz equation, $\nabla^2V + k^2V = 0$, where $k$ is the wavenumber, can be written in prolate spheroidal coordinates as
\begin{equation}
\label{x036}
\left(\frac{\partial}{\partial\eta}\left(1 - \eta^2\right)\frac{\partial}{\partial\eta} + \frac{\partial}{\partial\xi}\left(\xi^2 - 1\right)\frac{\partial}{\partial\xi} + \frac{\xi^2 - \eta^2}{\left(1 - \eta^2\right)\left(\xi^2 - 1\right)}\frac{\partial^2}{\partial\phi^2} + c^2\left(\xi^2 - \eta^2\right)\right)V = 0,
\end{equation}
where $c = ka$.
Applying the method of separation of variables yields three uncoupled ordinary differential equations, one for each coordinate:
\begin{equation}
\label{x004}
\frac{\partial}{\partial{}\eta}\left(\left(1 - \eta^2\right)\frac{\partial}{\partial{}\eta}S_{mn}\left(c, \eta\right)\right) + \left(\lambda_{mn} - c^2\eta^2 - \frac{m^2}{1 - \eta^2}\right)S_{mn}\left(c, \eta\right) = 0,
\end{equation}
\begin{equation}
\label{x017}
\frac{\partial}{\partial{}\xi}\left(\left(\xi^2 - 1\right)\frac{\partial}{\partial{}\xi}R_{mn}\left(c, \xi\right)\right) - \left(\lambda_{mn} - c^2\xi^2 + \frac{m^2}{\xi^2 - 1}\right)R_{mn}\left(c, \xi\right) = 0,
\end{equation}
\begin{equation}
\label{x018}
\frac{\partial^2}{{\partial\phi}^2}\Phi_m\left(\phi\right) + m^2\Phi_m\left(\phi\right) = 0,
\end{equation}
where $m = 0, 1, \ldots$ and $n = m, m + 1, \ldots$.
While Eq.\ (\ref{x018}) is easily solved, Eqs.\ (\ref{x004}) and (\ref{x017}) are much more complicated.
The solutions to Eq.\ (\ref{x004}) are called the prolate spheroidal angle functions, and the solutions to Eq.\ (\ref{x017}) are called the prolate spheroidal radial functions.
Collectively, they are called the prolate spheroidal wave functions.
Any solution to Eq.\ (\ref{x036}) can be written as
\begin{equation}
\label{pro_sum}
V = \sum_{m = 0}^\infty\sum_{n = m}^\infty{}S_{mn}\left(c, \eta\right)\left(A_{mn}R_{mn}^{\left(1\right)}\left(c, \xi\right) + B_{mn}R_{mn}^{\left(3\right)}\left(c, \xi\right)\right)\cos\left(m\phi\right),
\end{equation}
where the expansion coefficients, $A_{mn}$ and $B_{mn}$, depend on the problem being solved.

The expressions arising in the oblate case are very similar to (and sometimes exactly the same as) those arising in the prolate case.
In many cases, simply letting $c, \xi \rightarrow -ic, i\xi$ provides a transformation from the prolate case to the oblate case \cite{flammer2005}.
Indeed, the preceeding paragraphs and equations for the prolate case can be transformed into those for the oblate case by using this transformation.
The oblate spheroidal coordinate system is shown in Figure \ref{y001}.

\section{Spheroidal Wave Functions}

\subsection{Prolate Spheroidal Wave Functions}

\begin{figure}[t]
	\centering
	\includegraphics[height=2.4in]{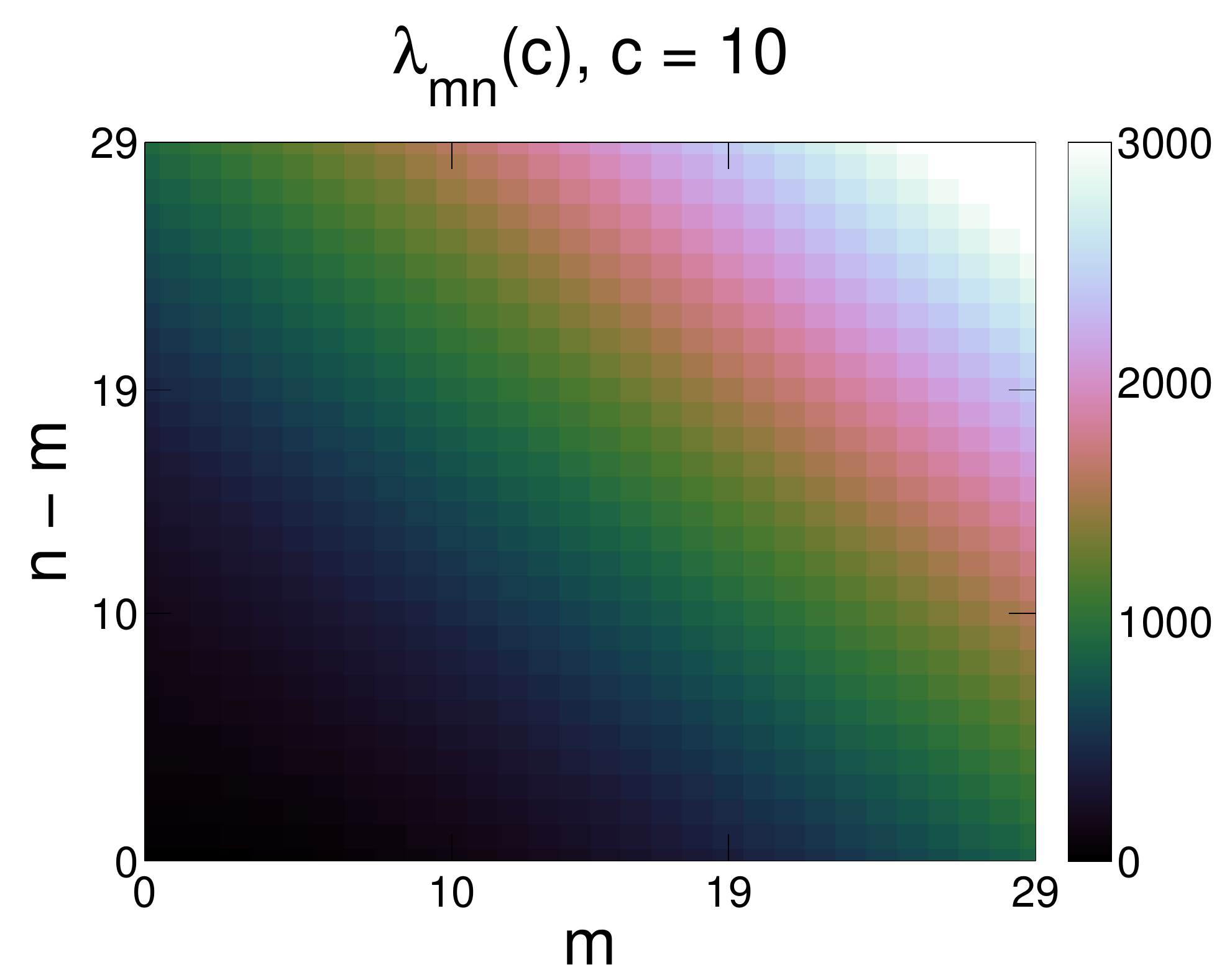}
	\includegraphics[height=2.4in]{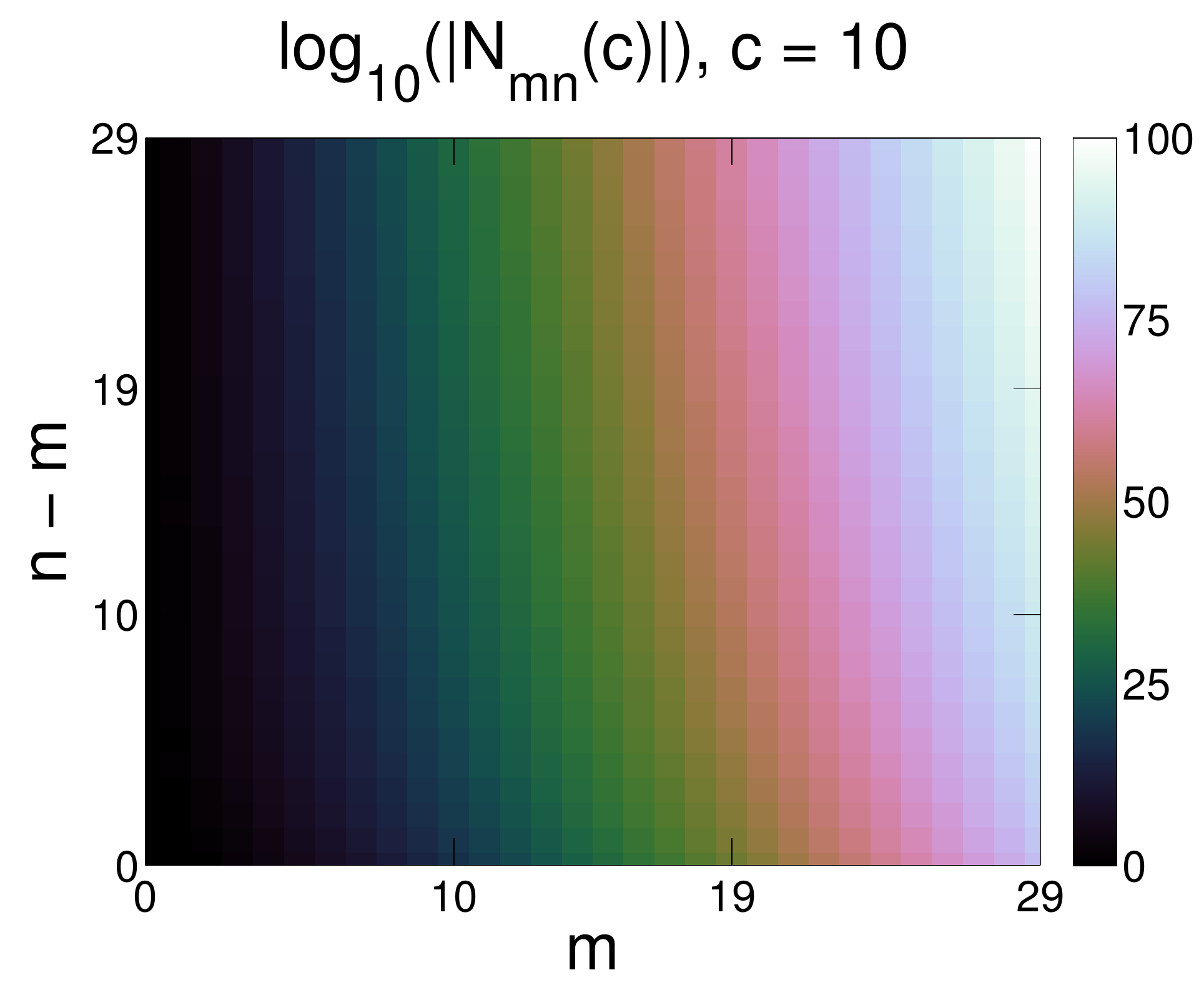}\\
	\includegraphics[height=2.4in]{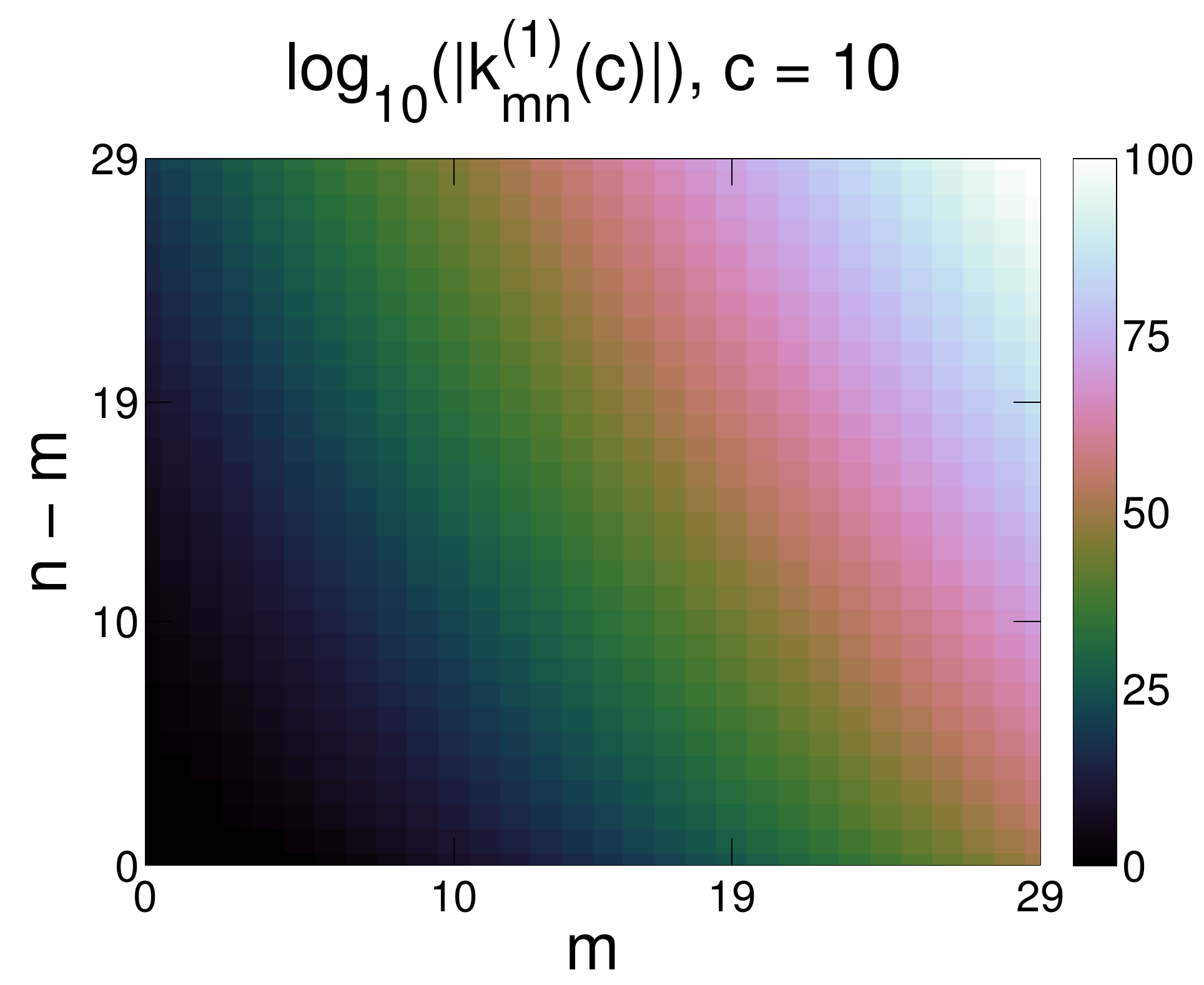}
	\includegraphics[height=2.4in]{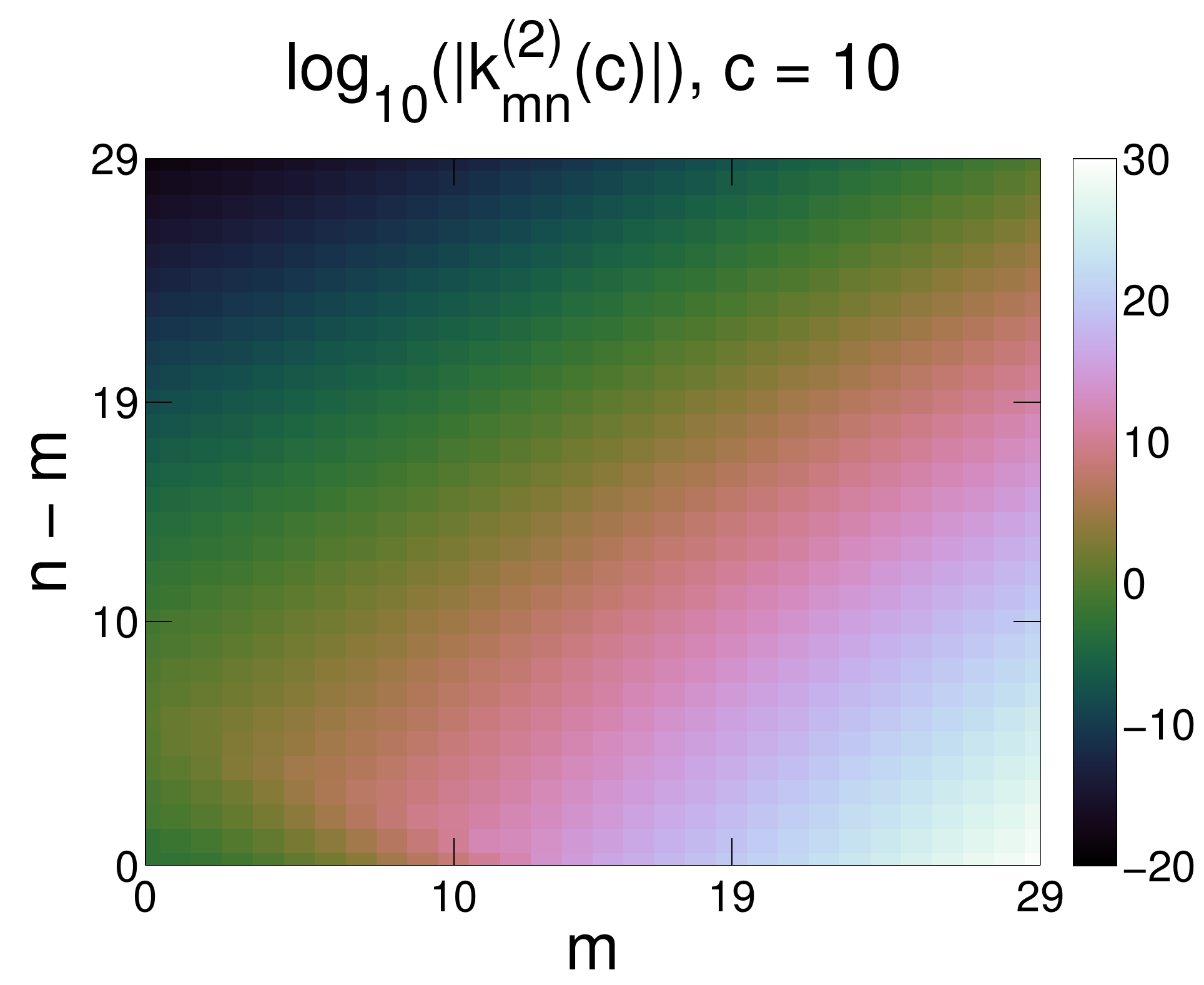}
	\caption{Characteristic and other special values for the prolate spheroidal wave functions for $c = 10$, $m = 0, 1, \ldots, 29$, and $n = m, m + 1, \ldots, m + 29$.}
	\label{pro_coefficients1}
\end{figure}

\begin{figure}[t]
	\centering
	\includegraphics[height=4.0in]{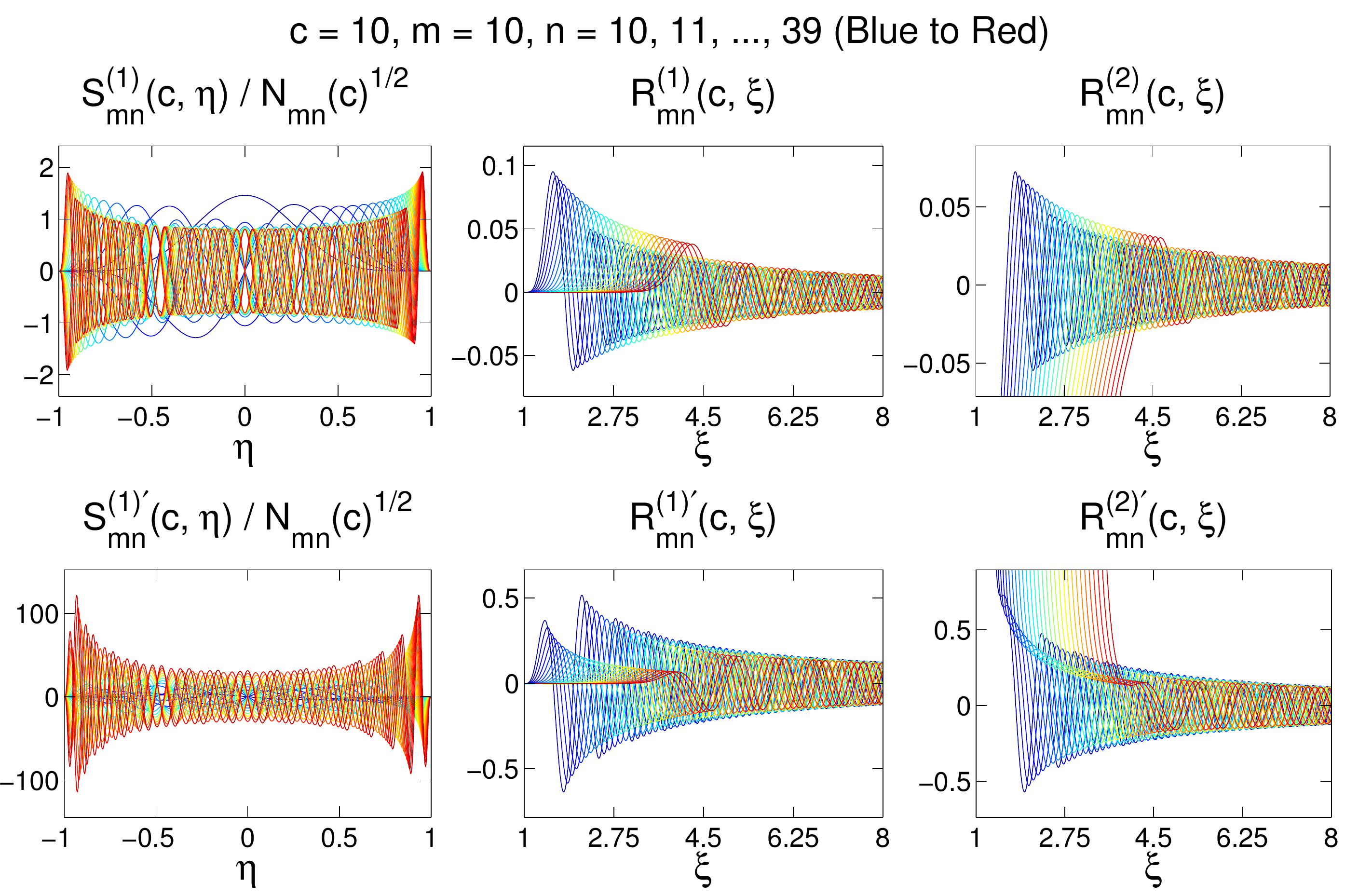}
	\caption{The prolate spheroidal wave functions and their derivatives for $c = 10$, $m = 10$, and $n = 10, 11, \ldots, 39$.}
	\label{pro_awesome3}
\end{figure}

\subsubsection{Angle Functions}

The prolate spheroidal angle functions of the first and second kinds can be written in terms of the associated Legendre polynomials of the first and second kinds, respectively:
\begin{equation}
\label{x005}
S_{mn}^{\left(1\right)}\left(c, \eta\right) = \psum_{r = 0, 1}^\infty{}d_r^{mn}\left(c\right)P_{m + r}^m\left(\eta\right),
\end{equation}
\begin{equation}
S_{mn}^{\left(2\right)}\left(c, \eta\right) = \psum_{r = -\infty}^\infty{}d_r^{mn}\left(c\right)Q_{m + r}^m\left(\eta\right),
\end{equation}
where the prime over the sum means that only the even terms are included when $n - m = \text{even}$ and only the odd terms are included when $n - m = \text{odd}$.
The angle functions of the first kind are orthogonal over $\left[-1, 1\right]$:
\begin{equation}
\int_{-1}^1S_{mn}^{\left(1\right)}\left(c, \eta\right)S_{mn^\prime}^{\left(1\right)}\left(c, \eta\right)d\eta = \delta_{nn^\prime}N_{mn}\left(c\right),
\end{equation}
where
\begin{equation}
\delta_{nn^\prime} = \left\{\begin{array}{c}
1,\quad{}n = n^\prime\\
0,\quad{}n \neq n^\prime
\end{array}\right.,\quad{}N_{mn}\left(c\right) = 2\psum_{r = 0, 1}^\infty{}d_r^{mn}\left(c\right)^2\frac{\left(2m + r\right)!}{\left(2m + 2r + 1\right)r!}.
\end{equation}

The angle functions of the first kind can also be written as a power series.
This can be done with the help of the hypergeometric function.
\noindent{}When $n - m = \text{even}$,
\begin{equation}
\label{x000}
S_{mn}^{\left(1\right)}\left(c, \eta\right) = \sum_{r = 0}^\infty{}d_{2r}^{mn}\left(c\right)P_{m + 2r}^m\left(\eta\right).
\end{equation}
The associated Legendre polynomials of the first kind can be written as
\begin{equation}
P_n^m\left(x\right) = \frac{\left(-1\right)^m\left(n + m\right)!}{2^mm!\left(n - m\right)!}\left(1 - x^2\right)^{m / 2}F\left(m - n, n + m + 1; m + 1; \frac{1 - x}{2}\right),
\end{equation}
where
\begin{equation}
F\left(\alpha, \beta; \gamma; x\right) = \sum_{k = 0}^\infty\frac{\left(\alpha\right)_k\left(\beta\right)_k}{k!\left(\gamma\right)_k}x^k
\end{equation}
is the hypergeometric function and
\begin{equation}
\left(a\right)_0 = 1,\quad\left(a\right)_k = a\left(a + 1\right)\ldots\left(a + k - 1\right)
\end{equation}
is the rising Pochhammer symbol.
Using this, the associated Legendre polynomials of the first kind in Eq.\ (\ref{x000}) can be written as
\begin{equation}
P_{m + 2r}^m\left(x\right) = \frac{\left(-1\right)^m\left(2m + 2r\right)!}{2^mm!\left(2r\right)!}\left(1 - x^2\right)^{m / 2}F\left(-2r, 2m + 2r + 1; m + 1; \frac{1 - x}{2}\right).
\end{equation}
Using the following identity for the hypergeometric function, we can rearrange the previous expression slightly:
\begin{equation}
F\left(\alpha, \beta; \frac{\alpha + \beta + 1}{2}; x\right) = F\left(\frac{\alpha}{2}, \frac{\beta}{2}; \frac{\alpha + \beta + 1}{2}; 4x\left(1 - x\right)\right),
\end{equation}
\begin{equation}
P_{m + 2r}^m\left(x\right) = \frac{\left(-1\right)^m\left(2m + 2r\right)!}{2^mm!\left(2r\right)!}\left(1 - x^2\right)^{m / 2}F\left(-r, m + r + \frac{1}{2}; m + 1; 1 - x^2\right),
\end{equation}
\begin{equation}
\label{x001}
P_{m + 2r}^m\left(x\right) = \frac{\left(-1\right)^m\left(2m + 2r\right)!}{2^mm!\left(2r\right)!}\left(1 - x^2\right)^{m / 2}\sum_{k = 0}^\infty\frac{\left(-r\right)_k\left(m + r + \frac{1}{2}\right)_k}{k!\left(m + 1\right)_k}\left(1 - x^2\right)^k.
\end{equation}
Plugging this into Eq.\ (\ref{x000}), we have
\begin{equation}
S_{mn}^{\left(1\right)}\left(c, \eta\right) = \sum_{r = 0}^\infty{}d_{2r}^{mn}\left(c\right)\frac{\left(-1\right)^m\left(2m + 2r\right)!}{2^mm!\left(2r\right)!}\left(1 - \eta^2\right)^{m / 2}\sum_{k = 0}^\infty\frac{\left(-r\right)_k\left(m + r + \frac{1}{2}\right)_k}{k!\left(m + 1\right)_k}\left(1 - \eta^2\right)^k.
\end{equation}
Rearranging,
\begin{equation}
S_{mn}^{\left(1\right)}\left(c, \eta\right) = \left(-1\right)^m\left(1 - \eta^2\right)^{m / 2}\sum_{k = 0}^\infty{}c_{2k}^{mn}\left(c\right)\left(1 - \eta^2\right)^k,
\end{equation}
where
\begin{equation}
c_{2k}^{mn}\left(c\right) = \frac{1}{2^m\left(m + k\right)!k!}\psum_{r = 2k}^\infty{}d_r^{mn}\left(c\right)\frac{\left(2m + r\right)!}{r!}\left(-\frac{r}{2}\right)_k\left(m + \frac{r}{2} + \frac{1}{2}\right)_k.
\end{equation}
\noindent{}When $n - m = \text{odd}$,
\begin{equation}
\label{x002}
S_{mn}^{\left(1\right)}\left(c, \eta\right) = \sum_{r = 0}^\infty{}d_{2r + 1}^{mn}\left(c\right)P_{m + 2r + 1}^m\left(\eta\right).
\end{equation}
Recall the following identity for the associated Legendre polynomials of the first kind:
\begin{equation}
\left(n - m + 1\right)P_{n + 1}^m\left(x\right) = \left(n + 1\right)xP_n^m\left(x\right) - \left(1 - x^2\right)P_n^{m\prime}\left(x\right).
\end{equation}
Setting $n = m + 2r$, plugging in Eq.\ (\ref{x001}), and rearranging,
\begin{equation}
P_{m + 2r + 1}^m\left(x\right) = \frac{\left(-1\right)^m\left(2m + 2r + 1\right)!}{2^mm!\left(2r + 1\right)!}x\left(1 - x^2\right)^{m / 2}\sum_{k = 0}^\infty\frac{\left(-r\right)_k\left(m + r + \frac{3}{2}\right)_k}{k!\left(m + 1\right)_k}\left(1 - x^2\right)^k.
\end{equation}
Finally, plugging this into Eq.\ (\ref{x002}),
\begin{equation}
\begin{array}{c}
\displaystyle{S_{mn}^{\left(1\right)}\left(c, \eta\right) = \sum_{r = 0}^\infty{}d_{2r + 1}^{mn}\left(c\right)\frac{\left(-1\right)^m\left(2m + 2r + 1\right)!}{2^mm!\left(2r + 1\right)!}\eta\left(1 - \eta^2\right)^{m / 2}\times}\\
[-0.1in]\\
\displaystyle{\sum_{k = 0}^\infty\frac{\left(-r\right)_k\left(m + r + \frac{3}{2}\right)_k}{k!\left(m + 1\right)_k}\left(1 - \eta^2\right)^k}.
\end{array}
\end{equation}
Rearranging,
\begin{equation}
S_{mn}^{\left(1\right)}\left(c, \eta\right) = \left(-1\right)^m\eta\left(1 - \eta^2\right)^{m / 2}\sum_{k = 0}^\infty{}c_{2k}^{mn}\left(c\right)\left(1 - \eta^2\right)^k,
\end{equation}
where
\begin{equation}
c_{2k}^{mn}\left(c\right) = \frac{1}{2^m\left(m + k\right)!k!}\psum_{r = 2k + 1}^\infty{}d_{r}^{mn}\left(c\right)\frac{\left(2m + r\right)!}{r!}\left(-\frac{r - 1}{2}\right)_k\left(m + \frac{r}{2} + 1\right)_k.
\end{equation}

\subsubsection{Radial Functions}

The prolate spheroidal radial functions of the first and second kinds can be written in terms of the spherical Bessel and Neumann functions, respectively:
\begin{equation}
R_{mn}^{\left(1\right)}\left(c, \xi\right) = {F_{mn}\left(c\right)}^{-1}\left(1 - \frac{1}{\xi^2}\right)^{m / 2}\psum_{r = 0, 1}^\infty{}\left(-1\right)^{\left(r - \left(n - m\right)\right) / 2}d_r^{mn}\left(c\right)\frac{\left(2m + r\right)!}{r!}j_{m + r}\left(c\xi\right),
\end{equation}
\begin{equation}
R_{mn}^{\left(2\right)}\left(c, \xi\right) = {F_{mn}\left(c\right)}^{-1}\left(1 - \frac{1}{\xi^2}\right)^{m / 2}\psum_{r = 0, 1}^\infty{}\left(-1\right)^{\left(r - \left(n - m\right)\right) / 2}d_r^{mn}\left(c\right)\frac{\left(2m + r\right)!}{r!}y_{m + r}\left(c\xi\right),
\end{equation}
where
\begin{equation}
F_{mn}\left(c\right) = \psum_{r = 0, 1}^\infty{}d_r^{mn}\left(c\right)\frac{\left(2m + r\right)!}{r!}.
\end{equation}
The radial functions of the third and fourth kinds are linear combinations of those of the first and second kinds:
\begin{equation}
R_{mn}^{\left(3\right)}\left(c, \xi\right) = R_{mn}^{\left(1\right)}\left(c, \xi\right) + iR_{mn}^{\left(2\right)}\left(c, \xi\right),
\end{equation}
\begin{equation}
R_{mn}^{\left(4\right)}\left(c, \xi\right) = R_{mn}^{\left(1\right)}\left(c, \xi\right) - iR_{mn}^{\left(2\right)}\left(c, \xi\right).
\end{equation}
The Wronskian of the radial functions of the first and second kinds is given by
\begin{equation}
W_{mn}\left(c, \xi\right) = R_{mn}^{\left(1\right)}\left(c, \xi\right)\frac{\partial}{\partial\xi}R_{mn}^{\left(2\right)}\left(c, \xi\right) - \frac{\partial}{\partial\xi}R_{mn}^{\left(1\right)}\left(c, \xi\right)R_{mn}^{\left(2\right)}\left(c, \xi\right) = \frac{1}{c\left(\xi^2 - 1\right)}
\end{equation}
and is useful for validating computed values of these functions.

The radial functions are related to the angle functions by
\begin{equation}
S_{mn}^{\left(1\right)}\left(c, z\right) = k_{mn}^{\left(1\right)}\left(c\right)R_{mn}^{\left(1\right)}\left(c, z\right),
\end{equation}
\begin{equation}
S_{mn}^{\left(2\right)}\left(c, z\right) = k_{mn}^{\left(2\right)}\left(c\right)R_{mn}^{\left(2\right)}\left(c, z\right),
\end{equation}
where $k_{mn}^{\left(1\right)}\left(c\right)$ is given by
\begin{equation}
k_{mn}^{\left(1\right)}\left(c\right) = \cfrac{\left(2m + 1\right)\left(m + n\right)!F_{mn}\left(c\right)}{2^{m + n}d_0^{mn}\left(c\right)c^mm!\left(\cfrac{n - m}{2}\right)!\left(\cfrac{m + n}{2}\right)!},\quad{}n - m = \text{even},
\end{equation}
\begin{equation}
k_{mn}^{\left(1\right)}\left(c\right) = \cfrac{\left(2m + 3\right)\left(m + n + 1\right)!F_{mn}\left(c\right)}{2^{m + n}d_1^{mn}\left(c\right)c^{m + 1}m!\left(\cfrac{n - m - 1}{2}\right)!\left(\cfrac{m + n + 1}{2}\right)!},\quad{}n - m = \text{odd},
\end{equation}
and $k_{mn}^{\left(2\right)}\left(c\right)$ is given by
\begin{equation}
k_{mn}^{\left(2\right)}\left(c\right) = \cfrac{2^{n - m}\left(2m\right)!\left(\cfrac{n - m}{2}\right)!\left(\cfrac{m + n}{2}\right)!d_{-2m}^{mn}\left(c\right)F_{mn}\left(c\right)}{\left(2m - 1\right)m!\left(m + n\right)!c^{m - 1}},\quad{}n - m = \text{even},
\end{equation}
\begin{equation}
k_{mn}^{\left(2\right)}\left(c\right) = -\cfrac{2^{n - m}\left(2m\right)!\left(\cfrac{n - m - 1}{2}\right)!\left(\cfrac{m + n + 1}{2}\right)!d_{-2m + 1}^{mn}\left(c\right)F_{mn}\left(c\right)}{\left(2m - 3\right)\left(2m - 1\right)m!\left(m + n + 1\right)!c^{m - 2}},\quad{}n - m = \text{odd}.
\end{equation}
The expression for the radial functions of the second kind using the spherical Neumann functions converges very slowly for values of $\xi$ near $1$ and is, therefore, inaccurate in these cases.
While the expression for the radial functions of the first kind is accurate for all values of $\xi$, having a second method can be used as a check on the first method.
Thus, these relationships can be used to construct secondary methods for computing these functions.
For the radial functions of the first kind,
\begin{equation}
R_{mn}^{\left(1\right)}\left(c, \xi\right) = {k_{mn}^{\left(1\right)}\left(c\right)}^{-1}S_{mn}^{\left(1\right)}\left(c, \xi\right),
\end{equation}
\begin{equation}
R_{mn}^{\left(1\right)}\left(c, \xi\right) = {k_{mn}^{\left(1\right)}\left(c\right)}^{-1}\psum_{r = 0, 1}^\infty{}d_r^{mn}\left(c\right)P_{m + r}^m\left(\xi\right).
\end{equation}
Similar to the angle functions of the first kind, this expression can be written as a power series with the help of the hypergeometric function.
Because the argument is $\xi \ge 1$ as opposed to $\left|\eta\right| \le 1$, the relationship between the associated Legendre polynomials of the first kind and the hypergeometric function is slightly different.
In particular, there is no factor of $\left(-1\right)^m$:
\begin{equation}
P_n^m\left(x\right) = \frac{\left(n + m\right)!}{2^mm!\left(n - m\right)!}\left(x^2 - 1\right)^{m / 2}F\left(m - n, n + m + 1; m + 1; \frac{1 - x}{2}\right).
\end{equation}
Following a procedure similar to the one followed for the angle functions of the first kind, we have
\begin{equation}
R_{mn}^{\left(1\right)}\left(c, \xi\right) = {k_{mn}^{\left(1\right)}\left(c\right)}^{-1}\left(\xi^2 - 1\right)^{m / 2}\sum_{k = 0}^\infty\left(-1\right)^kc_{2k}^{mn}\left(c\right)\left(\xi^2 - 1\right)^k,\quad{}n - m = \text{even},
\end{equation}
\begin{equation}
R_{mn}^{\left(1\right)}\left(c, \xi\right) = {k_{mn}^{\left(1\right)}\left(c\right)}^{-1}\xi\left(\xi^2 - 1\right)^{m / 2}\sum_{k = 0}^\infty\left(-1\right)^kc_{2k}^{mn}\left(c\right)\left(\xi^2 - 1\right)^k,\quad{}n - m = \text{odd},
\end{equation}
where $c_{2k}^{mn}\left(c\right)$ is the same as before.
For the radial functions of the second kind,
\begin{equation}
R_{mn}^{\left(2\right)}\left(c, \xi\right) = {k_{mn}^{\left(2\right)}\left(c\right)}^{-1}S_{mn}^{\left(2\right)}\left(c, \xi\right),
\end{equation}
\begin{equation}
\label{x003}
R_{mn}^{\left(2\right)}\left(c, \xi\right) = {k_{mn}^{\left(2\right)}\left(c\right)}^{-1}\psum_{r = -\infty}^\infty{}d_r^{mn}\left(c\right)Q_{m + r}^m\left(\xi\right).
\end{equation}

\subsubsection{Calculating the Characteristic Value and Expansion Coefficients}

\label{z001}

All of the expressions introduced in the previous sections require, either directly or indirectly, the characteristic value, $\lambda_{mn}\left(c\right)$, and expansion coefficients, $d_r^{mn}\left(c\right)$.
Below, we derive expressions for computing them.

There are several methods for computing the characterstic value, but here, we use a combination of two: method (1) involves solving for the eigenvalues of a tridiagonal matrix \cite{hodge1970}; and method (2) involves solving for the roots of a transcendental equation \cite{flammer2005}.
To begin, method (1) is used to compute an approximate value for the characteristic value.
Then, method (2) is used to compute a more accurate value for the characteristic value using the approximate value computed by method (1) as a starting point.
This procedure is similar to the one used in \cite{falloon2003}.
In our software, we use double precision for method (1) and arbitrary precision for method (2).

Both methods rely on the following recurrence relation, which can be obtained by plugging Eq.\ (\ref{x005}) into Eq.\ (\ref{x004}):
\begin{equation}
\label{x009}
\alpha_rd_{r + 2}^{mn}\left(c\right) + \left(\beta_r - \lambda_{mn}\left(c\right)\right)d_r^{mn}\left(c\right) + \gamma_rd_{r - 2}^{mn}\left(c\right) = 0,
\end{equation}
where
\begin{equation}
\alpha_r = \frac{\left(2m + r + 2\right)\left(2m + r + 1\right)}{\left(2m + 2r + 5\right)\left(2m + 2r + 3\right)}c^2,
\end{equation}
\begin{equation}
\beta_r = \left(m + r\right)\left(m + r + 1\right) + \frac{2\left(m + r\right)\left(m + r + 1\right) - 2m^2 - 1}{\left(2m + 2r - 1\right)\left(2m + 2r + 3\right)}c^2,
\end{equation}
\begin{equation}
\gamma_r = \frac{r\left(r - 1\right)}{\left(2m + 2r - 3\right)\left(2m + 2r - 1\right)}c^2.
\end{equation}

For method (1), this recurrence relation is rearranged slightly:
\begin{equation}
\alpha_rd_{r + 2}^{mn}\left(c\right) + \beta_rd_r^{mn}\left(c\right) + \gamma_rd_{r - 2}^{mn}\left(c\right) = \lambda_{mn}\left(c\right)d_r^{mn}\left(c\right).
\end{equation}
In matrix form,
\begin{equation}
\left[\begin{array}{ccccc}
\beta_0&\alpha_0&&\\
\gamma_2&\beta_2&\alpha_2&\\
&\gamma_4&\beta_4&\alpha_4\\
&&&\ddots
\end{array}\right]\left[\begin{array}{c}
d_0^{mn}\left(c\right)\\
d_2^{mn}\left(c\right)\\
d_4^{mn}\left(c\right)\\
\vdots
\end{array}\right] = \lambda_{mn}\left(c\right)\left[\begin{array}{c}
d_0^{mn}\left(c\right)\\
d_2^{mn}\left(c\right)\\
d_4^{mn}\left(c\right)\\
\vdots
\end{array}\right],\quad{}n - m = \text{even},
\end{equation}
\begin{equation}
\left[\begin{array}{ccccc}
\beta_1&\alpha_1&&\\
\gamma_3&\beta_3&\alpha_3&\\
&\gamma_5&\beta_5&\alpha_5\\
&&&\ddots
\end{array}\right]\left[\begin{array}{c}
d_1^{mn}\left(c\right)\\
d_3^{mn}\left(c\right)\\
d_5^{mn}\left(c\right)\\
\vdots
\end{array}\right] = \lambda_{mn}\left(c\right)\left[\begin{array}{c}
d_1^{mn}\left(c\right)\\
d_3^{mn}\left(c\right)\\
d_5^{mn}\left(c\right)\\
\vdots
\end{array}\right],\quad{}n - m = \text{odd}.
\end{equation}
When $n - m = \text{even}$, the eigenvalues are $\lambda_{mn}\left(c\right)$ for $n = m, m + 2, m + 4, \ldots$, and when $n - m = \text{odd}$, the eigenvalues are $\lambda_{mn}\left(c\right)$ for $n = m + 1, m + 3, m + 5, \ldots$.
Thus, we can compute $\lambda_{mn}\left(c\right)$ by plugging these tridiagonal matrices into an eigenvalue solver.
In our software, we use the \verb#eig# function in MATLAB.

In method (2), the recurrence relation in Eq.\ (\ref{x009}) is divided through by $d_r^{mn}\left(c\right)$, which yields
\begin{equation}
\label{x010}
\alpha_r\frac{d_{r + 2}^{mn}\left(c\right)}{d_r^{mn}\left(c\right)} + \beta_r - \lambda_{mn}\left(c\right) + \gamma_r\frac{d_{r - 2}^{mn}\left(c\right)}{d_r^{mn}\left(c\right)} = 0.
\end{equation}
Setting
\begin{equation}
\label{x011}
N_r^m = -\alpha_{r - 2}\frac{d_r^{mn}\left(c\right)}{d_{r - 2}^{mn}\left(c\right)},\quad\gamma_r^m = \beta_r,\quad\beta_r^m = \gamma_r\alpha_{r - 2}
\end{equation}
allows us to write Eq.\ (\ref{x010}) as
\begin{equation}
-N_{r + 2}^m + \gamma_r^m - \lambda_{mn}\left(c\right) - \frac{\beta_r^m}{N_r^m} = 0.
\end{equation}
Rearranging one way leads to a continued fraction in decreasing $r$:
\begin{equation}
\label{x006}
N_r^m = \gamma_{r - 2}^m - \lambda_{mn}\left(c\right) - \frac{\beta_{r - 2}^m}{\gamma_{r - 4}^m - \lambda_{mn}\left(c\right) - }\frac{\beta_{r - 4}^m}{\gamma_{r - 6}^m - \lambda_{mn}\left(c\right) - }\cdots.
\end{equation}
Rearranging the other way leads to a continued fraction in increasing $r$:
\begin{equation}
\label{x007}
N_r^m = \frac{\beta_r^m}{\gamma_r^m - \lambda_{mn}\left(c\right) - }\frac{\beta_{r + 2}^m}{\gamma_{r + 2}^m - \lambda_{mn}\left(c\right) - }\cdots.
\end{equation}
The two expression for $N_r^m$ should be equal to each other.
Setting $r = n - m + 2$ in Eqs.\ (\ref{x006}) and (\ref{x007}),
\begin{equation}
U_1\left(\lambda_{mn}\left(c\right)\right) = \gamma_{n - m}^m - \lambda_{mn}\left(c\right) - \frac{\beta_{n - m}^m}{\gamma_{n - m - 2}^m - \lambda_{mn}\left(c\right) - }\frac{\beta_{n - m - 2}^m}{\gamma_{n - m - 4}^m - \lambda_{mn}\left(c\right) - }\cdots,
\end{equation}
\begin{equation}
U_2\left(\lambda_{mn}\left(c\right)\right) = -\frac{\beta_{n - m + 2}^m}{\gamma_{n - m + 2}^m - \lambda_{mn}\left(c\right) - }\frac{\beta_{n - m + 4}^m}{\gamma_{n - m + 4}^m - \lambda_{mn}\left(c\right) - }\cdots.
\end{equation}
Adding these together yields a transcendental equation in $\lambda_{mn}\left(c\right)$:
\begin{equation}
\label{x008}
U\left(\lambda_{mn}\left(c\right)\right) = U_1\left(\lambda_{mn}\left(c\right)\right) + U_2\left(\lambda_{mn}\left(c\right)\right) = 0.
\end{equation}
We can compute $\lambda_{mn}\left(c\right)$ by solving this transcendental equation.
In our software, we use the secant method.

Once $\lambda_{mn}\left(c\right)$ is known, Eqs.\ (\ref{x011}) and (\ref{x007}) can be used to compute $d_r^{mn}\left(c\right)$.
The expansion coefficients are unique up to a constant factor, though, so the following normalization scheme is used.
When $n - m = \text{even}$,
\begin{equation}
S_{mn}^{\left(1\right)}\left(c, 0\right) = P_n^m\left(0\right),
\end{equation}
\begin{equation}
\psum_{r = 0}^\infty{}d_r^{mn}\left(c\right)\cfrac{\left(-1\right)^{r / 2}\left(2m + r\right)!}{2^r\left(\cfrac{2m + r}{2}\right)!\left(\cfrac{r}{2}\right)!} = \cfrac{\left(-1\right)^{\left(n - m\right) / 2}\left(n + m\right)!}{2^{n - m}\left(\cfrac{n + m}{2}\right)!\left(\cfrac{n - m}{2}\right)!}.
\end{equation}
When $n - m = \text{odd}$,
\begin{equation}
S_{mn}^{\left(1\right)\prime}\left(c, 0\right) = P_n^{m\prime}\left(0\right),
\end{equation}
\begin{equation}
\psum_{r = 1}^\infty{}d_r^{mn}\left(c\right)\cfrac{\left(-1\right)^{\left(r - 1\right) / 2}\left(2m + r + 1\right)!}{2^r\left(\cfrac{2m + r + 1}{2}\right)!\left(\cfrac{r - 1}{2}\right)!} = \cfrac{\left(-1\right)^{\left(n - m - 1\right) / 2}\left(n + m + 1\right)!}{2^{n - m}\left(\cfrac{n + m + 1}{2}\right)!\left(\cfrac{n - m - 1}{2}\right)!}.
\end{equation}

To use Eq.\ (\ref{x003}), $d_r^{mn}\left(c\right)$ must be computed for negative $r$ as well.
To begin, Eq.\ (\ref{x009}) is rewritten as
\begin{equation}
A_{r + 2}^md_{r + 2}^{mn}\left(c\right) + B_r^md_r^{mn}\left(c\right) + C_{r - 2}^md_{r - 2}^{mn}\left(c\right) = 0,
\end{equation}
where
\begin{equation}
A_r^m = \alpha_{r - 2},\quad{}B_r^m = \beta_r - \lambda_{mn}\left(c\right),\quad{}C_r^m = \gamma_{r + 2}.
\end{equation}
Rearranging,
\begin{equation}
\frac{d_r^{mn}\left(c\right)}{d_{r + 2}^{mn}\left(c\right)} = -\cfrac{A_{r + 2}^m}{B_r^m + C_{r - 2}^m\cfrac{d_{r - 2}^{mn}\left(c\right)}{d_r^{mn}\left(c\right)}},
\end{equation}
which can be expanded as a continued fraction in decreasing $r$:
\begin{equation}
\frac{d_r^{mn}\left(c\right)}{d_{r + 2}^{mn}\left(c\right)} = -\frac{A_{r + 2}^m}{B_r^m - }\frac{C_{r - 2}^mA_r^m}{B_{r - 2}^m - }\frac{C_{r - 4}^mA_{r - 2}^m}{B_{r - 4}^m - }\cdots.
\end{equation}
Because $A_r^m = 0$ when $r = -2m$ or $r = -2m + 1$, this continued fraction ends:
\begin{equation}
\frac{d_r^{mn}\left(c\right)}{d_{r + 2}^{mn}\left(c\right)} = -\frac{A_{r + 2}^m}{B_r^m - }\frac{C_{r - 2}^mA_r^m}{B_{r - 2}^m - }\frac{C_{r - 4}^mA_{r - 2}^m}{B_{r - 4}^m - }\cdots\frac{A_{-2m + 2}^m}{B_{-2m}^m + C_{-2m - 2}^m}
\end{equation}
when $n - m = \text{even}$, and
\begin{equation}
\frac{d_r^{mn}\left(c\right)}{d_{r + 2}^{mn}\left(c\right)} = -\frac{A_{r + 2}^m}{B_r^m - }\frac{C_{r - 2}^mA_r^m}{B_{r - 2}^m - }\frac{C_{r - 4}^mA_{r - 2}^m}{B_{r - 4}^m - }\cdots\frac{A_{-2m + 3}^m}{B_{-2m + 1}^m + C_{-2m - 1}^m}
\end{equation}
when $n - m = \text{odd}$.
This also means that $d_r^{mn}\left(c\right) \rightarrow 0$ when $r \le -2m - 2$ for $n - m = \text{even}$ and $r \le -2m - 1$ for $n - m = \text{odd}$.
However, $Q_{m + r}^m\left(\xi\right) \rightarrow \infty$ in these cases, and $d_r^{mn}\left(c\right)Q_{m + r}^m\left(\xi\right) < \infty$:
\begin{equation}
d_r^{mn}\left(c\right)Q_{m + r}^m\left(\xi\right) = d_{r | \epsilon}^{mn}\left(c\right)P_{-r - m - 1}^m\left(\xi\right).
\end{equation}
The sum in Eq.\ (\ref{x003}) is, therefore, separated into two pieces:
\begin{equation}
R_{mn}^{\left(2\right)}\left(c, \xi\right) = {k_{mn}^{\left(2\right)}}^{-1}\left(\psum_{r = -\infty}^{-2m - 2, -2m - 1}{}d_{r | \epsilon}^{mn}\left(c\right)P_{-r - m - 1}^m\left(\xi\right) + \psum_{r = -2m, -2m + 1}^\infty{}d_r^{mn}\left(c\right)Q_{m + r}^m\left(\xi\right)\right),
\end{equation}
where $d_{-2m - 2 | \epsilon}^{mn}\left(c\right)$ and $d_{-2m - 1 | \epsilon}^{mn}\left(c\right)$ are computed using
\begin{equation}
\frac{d_{-2m - 2 | \epsilon}^{mn}\left(c\right)}{d_{-2m}^{mn}\left(c\right)} = \frac{c^2}{\left(2m - 1\right)\left(2m + 1\right)}\frac{1}{B_{-2m - 2}^m - }\frac{C_{-2m - 4}^mA_{-2m - 2}^m}{B_{-2m - 4}^m - }\frac{C_{-2m - 6}^mA_{-2m - 4}^m}{B_{-2m - 6}^m - }\cdots,
\end{equation}
\begin{equation}
\frac{d_{-2m - 1 | \epsilon}^{mn}\left(c\right)}{d_{-2m + 1}^{mn}\left(c\right)} = -\frac{c^2}{\left(2m - 1\right)\left(2m - 3\right)}\frac{1}{B_{-2m - 1}^m - }\frac{C_{-2m - 3}^mA_{-2m - 1}^m}{B_{-2m - 3}^m - }\frac{C_{-2m - 5}^mA_{-2m - 3}^m}{B_{-2m - 5}^m - }\cdots.
\end{equation}
For $r < -2m - 2$ when $n - m = \text{even}$ and $r < -2m - 1$ when $n - m = \text{odd}$, the remaining expansion coefficients can be computed using
\begin{equation}
\frac{d_{r | \epsilon}^{mn}\left(c\right)}{d_{r + 2 | \epsilon}^{mn}\left(c\right)} = -\frac{A_{r + 2}^m}{B_r^m - }\frac{C_{r - 2}^mA_r^m}{B_{r - 2}^m - }\frac{C_{r - 4}^mA_{r - 2}^m}{B_{r - 4}^m - }\cdots.
\end{equation}

\subsection{Oblate Spheroidal Wave Functions}

\begin{figure}[t]
	\centering
	\includegraphics[height=2.4in]{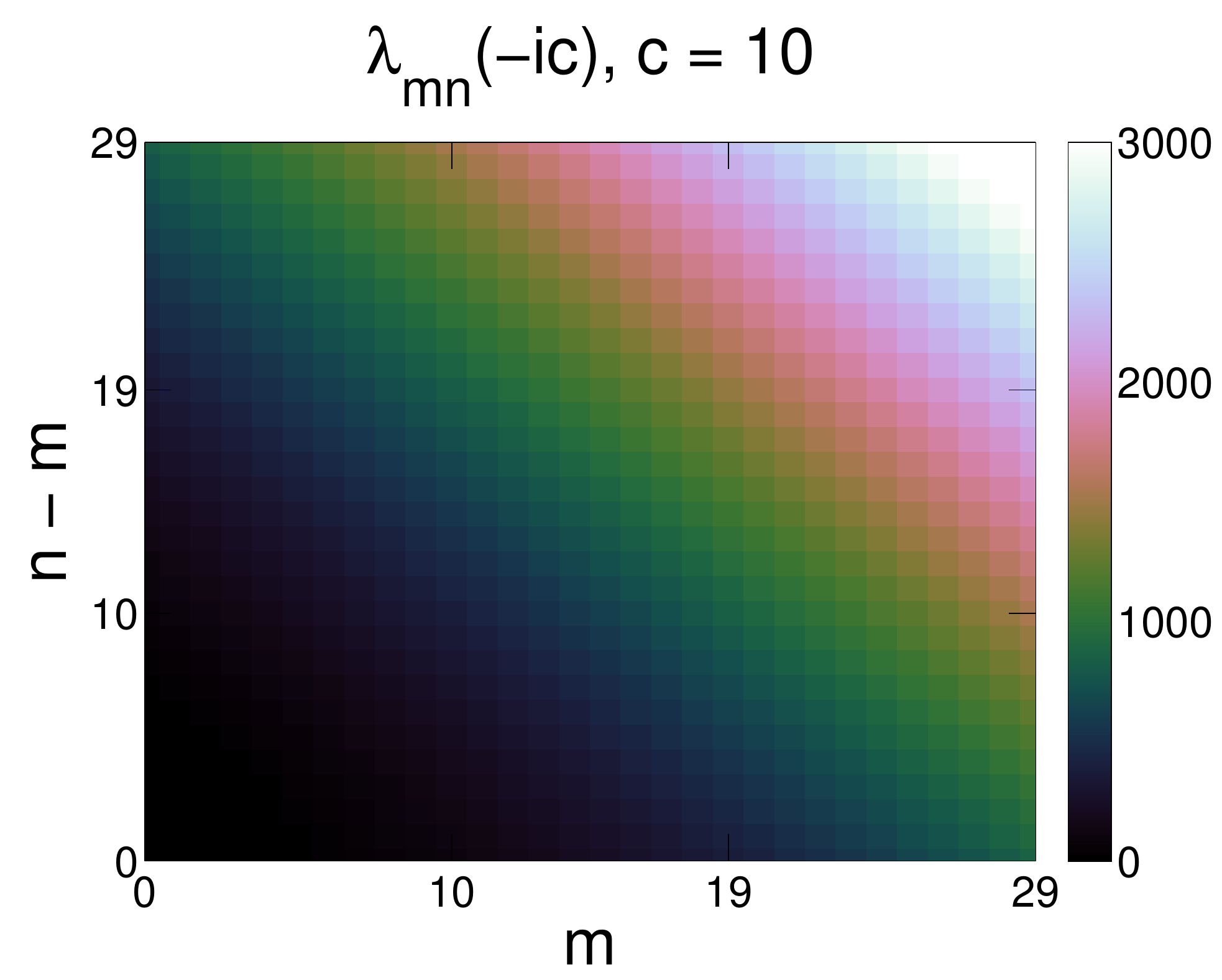}
	\includegraphics[height=2.4in]{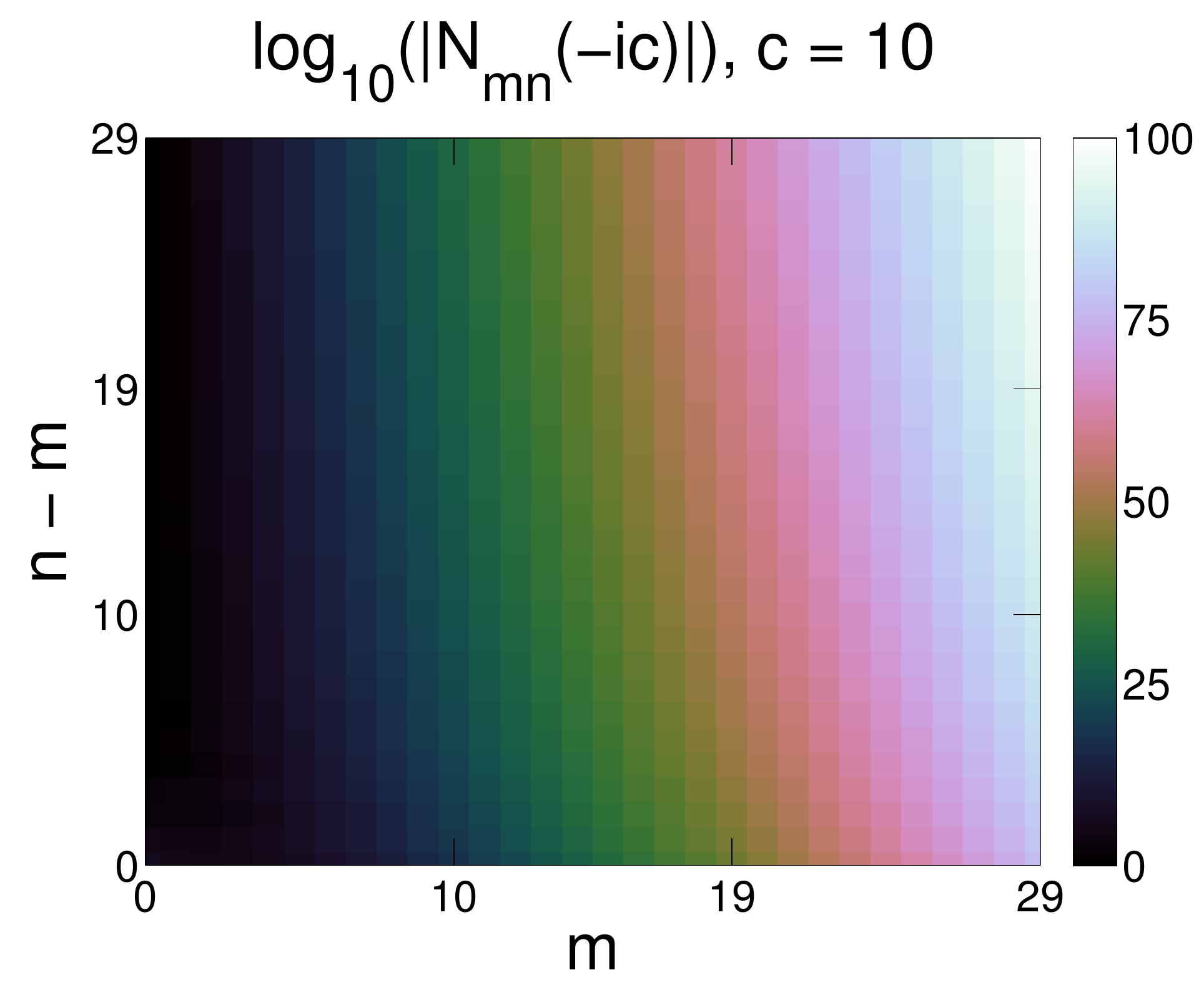}\\
	\includegraphics[height=2.4in]{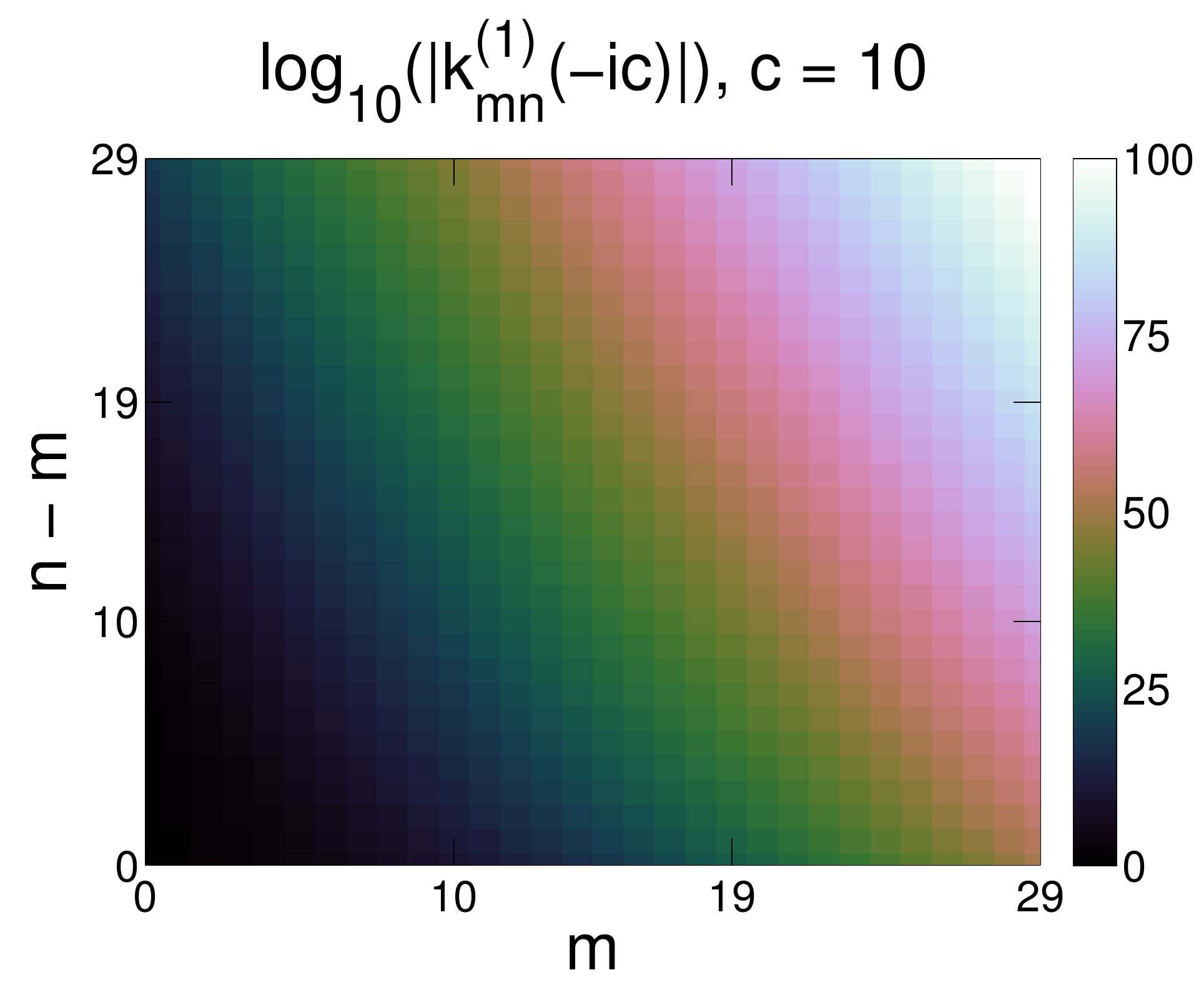}
	\includegraphics[height=2.4in]{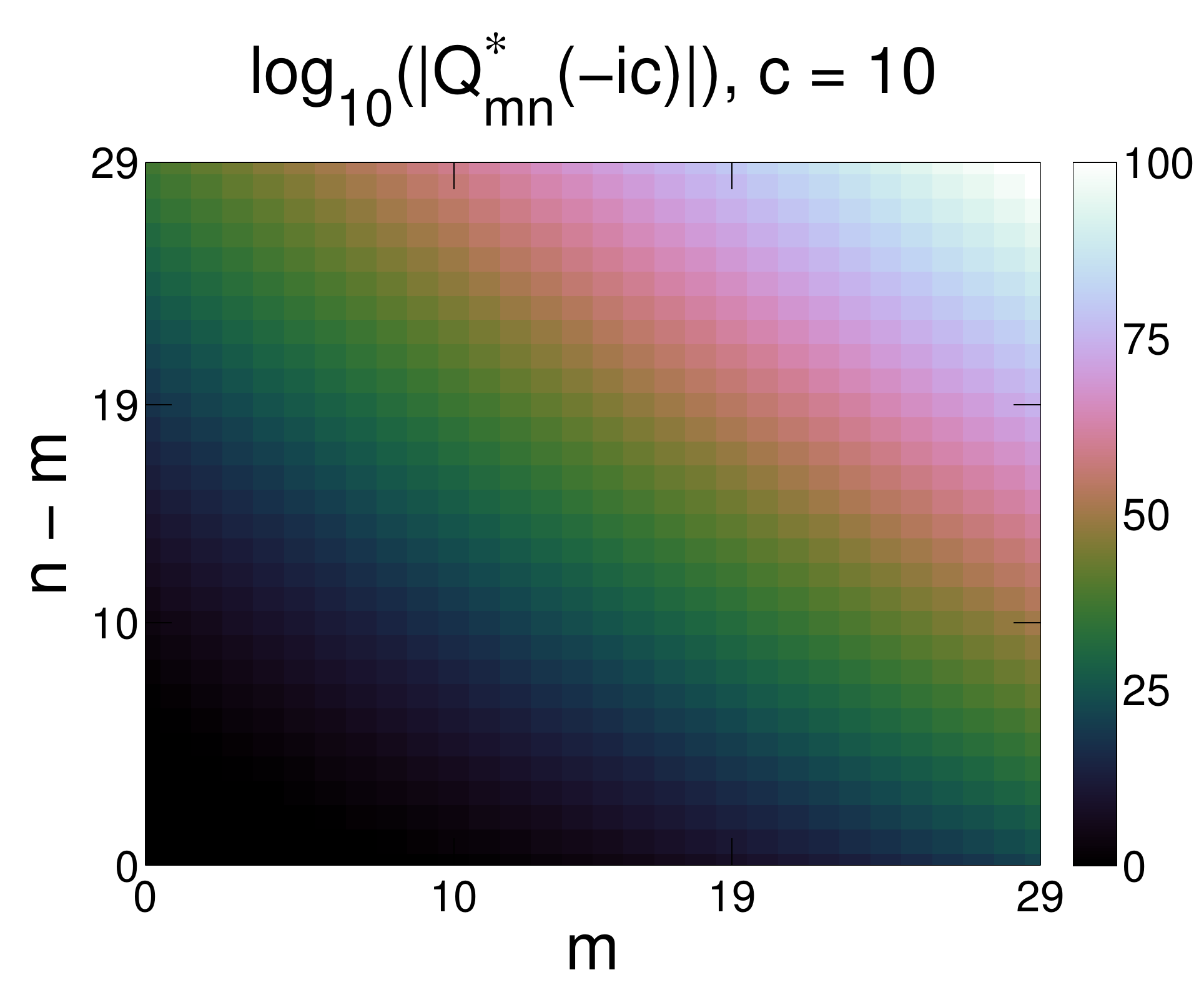}
	\caption{Characteristic and other special values for the oblate spheroidal wave functions for $c = 10$, $m = 0, 1, \ldots, 29$, and $n = m, m + 1, \ldots, m + 29$.}
	\label{obl_coefficients1}
\end{figure}

\begin{figure}[t]
	\centering
	\includegraphics[height=4.0in]{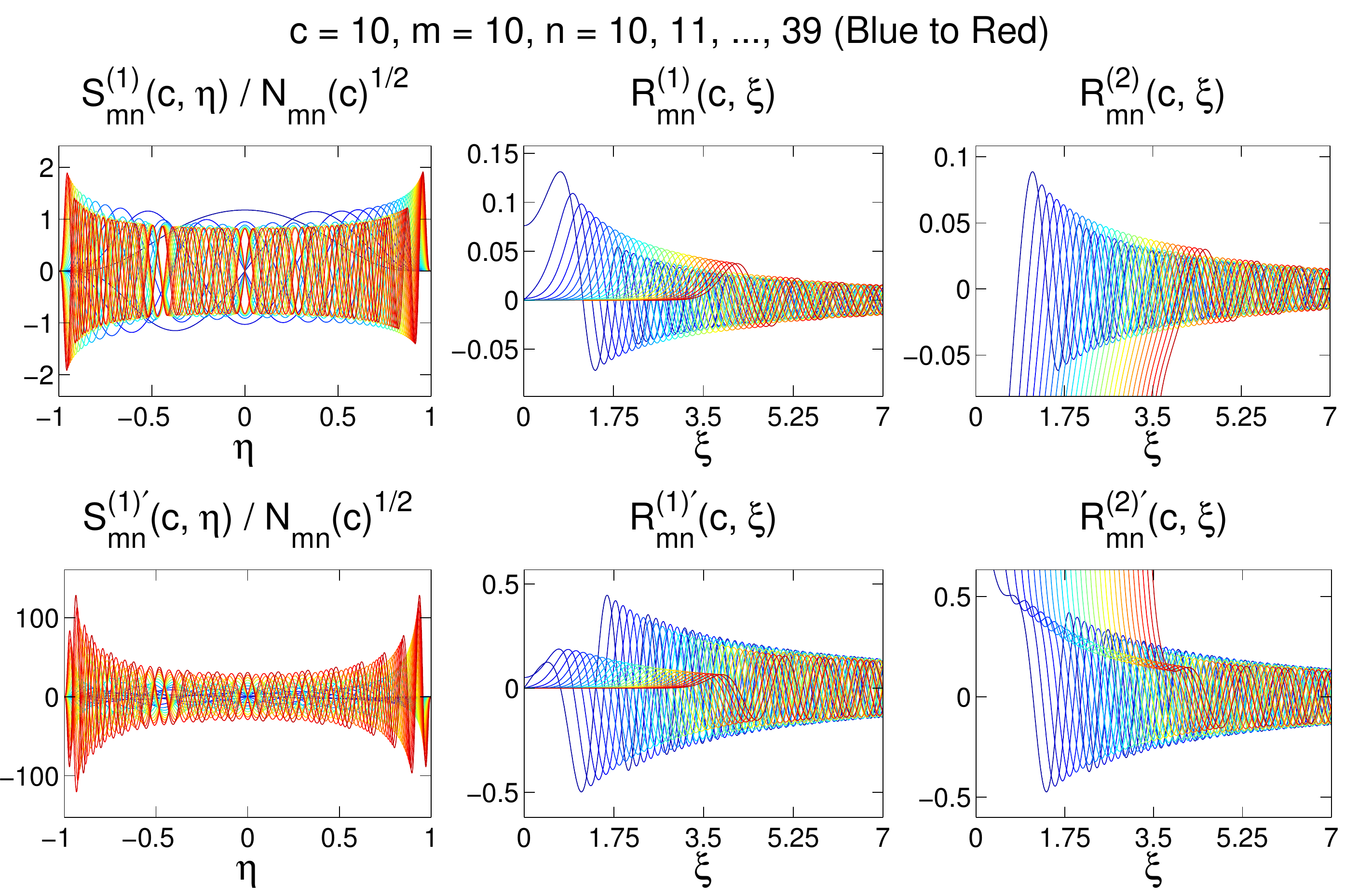}
	\caption{The oblate spheroidal wave functions and their derivatives for $c = 10$, $m = 10$, and $n = 10, 11, \ldots, 39$.}
	\label{obl_awesome3}
\end{figure}

\subsubsection{Angle Functions}

The expressions for the prolate spheroidal angle functions, including the ones written in terms of the associated Legendre polynomials as well as the ones expanded as power series, can be transformed into those for the oblate spheroidal angle functions by letting $c \rightarrow -ic$.

\subsubsection{Radial Functions}

The oblate spheroidal radial functions of the first and second kinds can be written in terms of the spherical Bessel and Neumann functions, respectively:
\begin{equation}
\begin{array}{c}
\displaystyle{R_{mn}^{\left(1\right)}\left(-ic, i\xi\right) = {F_{mn}\left(-ic\right)}^{-1}\left(1 + \frac{1}{\xi^2}\right)^{m / 2}\times}\\
[-0.1in]\\
\displaystyle{\psum_{r = 0, 1}^\infty{}\left(-1\right)^{\left(r - \left(n - m\right)\right) / 2}d_r^{mn}\left(-ic\right)\frac{\left(2m + r\right)!}{r!}j_{m + r}\left(c\xi\right),}
\end{array}
\end{equation}
\begin{equation}
\begin{array}{c}
\displaystyle{R_{mn}^{\left(2\right)}\left(-ic, i\xi\right) = {F_{mn}\left(-ic\right)}^{-1}\left(1 + \frac{1}{\xi^2}\right)^{m / 2}\times}\\
[-0.1in]\\
\displaystyle{\psum_{r = 0, 1}^\infty{}\left(-1\right)^{\left(r - \left(n - m\right)\right) / 2}d_r^{mn}\left(-ic\right)\frac{\left(2m + r\right)!}{r!}y_{m + r}\left(c\xi\right),}
\end{array}
\end{equation}
where
\begin{equation}
F_{mn}\left(-ic\right) = \psum_{r = 0, 1}^\infty{}d_r^{mn}\left(-ic\right)\frac{\left(2m + r\right)!}{r!}.
\end{equation}
The radial functions of the third and fourth kinds are linear combinations of those of the first and second kinds:
\begin{equation}
R_{mn}^{\left(3\right)}\left(-ic, i\xi\right) = R_{mn}^{\left(1\right)}\left(-ic, i\xi\right) + iR_{mn}^{\left(2\right)}\left(-ic, i\xi\right),
\end{equation}
\begin{equation}
R_{mn}^{\left(4\right)}\left(-ic, i\xi\right) = R_{mn}^{\left(1\right)}\left(-ic, i\xi\right) - iR_{mn}^{\left(2\right)}\left(-ic, i\xi\right).
\end{equation}
The Wronskian of the radial functions of the first and second kinds is given by
\begin{equation}
W_{mn}\left(-ic, i\xi\right) = R_{mn}^{\left(1\right)}\left(-ic, i\xi\right)\frac{\partial}{\partial\xi}R_{mn}^{\left(2\right)}\left(-ic, i\xi\right) - \frac{\partial}{\partial\xi}R_{mn}^{\left(1\right)}\left(-ic, i\xi\right)R_{mn}^{\left(2\right)}\left(-ic, i\xi\right) = \frac{1}{c\left(\xi^2 + 1\right)}
\end{equation}
and is useful for validating computed values of these functions.

The radial functions are related to the angle functions by
\begin{equation}
S_{mn}^{\left(1\right)}\left(-ic, iz\right) = k_{mn}^{\left(1\right)}\left(-ic\right)R_{mn}^{\left(1\right)}\left(-ic, iz\right),
\end{equation}
\begin{equation}
S_{mn}^{\left(2\right)}\left(-ic, iz\right) = k_{mn}^{\left(1\right)}\left(-ic\right)R_{mn}^{\left(2\right)}\left(-ic, iz\right),
\end{equation}
where $k_{mn}^{\left(1\right)}\left(-ic\right)$ and $k_{mn}^{\left(2\right)}\left(-ic\right)$ are given by the same expressions as in the prolate case, provided that $c \rightarrow -ic$.
The expression for the radial functions of the first kind using the spherical Bessel functions converges and is accurate for all values of $\xi$, except for $\xi = 0$, where the expression is undefined due to a divide by zero.
The expression for the radial functions of the second kind using the spherical Neumann functions converges very slowly for values of $\xi$ near $0$ and is, therefore, inaccurate in these cases.
Thus, these relationships can be used to construct secondary methods for computing these functions.
The same procedures used in the prolate case for doing so can also be used here, provided that $c, \xi \rightarrow -ic, i\xi$ where necessary.

A tertiary method for computing the radial functions of the second kind can also be constructed using a power series:
\begin{equation}
\label{x035}
R_{mn}^{\left(2\right)}\left(-ic, i\xi\right) = Q_{mn}^\ast\left(-ic\right)R_{mn}^{\left(1\right)}\left(-ic, i\xi\right)\left(\arctan\left(\xi\right) - \frac{\pi}{2}\right) + g_{mn}\left(-ic, i\xi\right),
\end{equation}
where
\begin{equation}
Q_{mn}^\ast\left(-ic\right) = \frac{\left(i^{-m}k_{mn}^{\left(1\right)}\left(-ic\right)\right)^2}{c}\sum_{r = 0}^m\alpha_{r}^{mn}\left(-ic\right)\frac{\left(2m - 2r\right)!}{r!\left(2^{m - r}\left(m - r\right)!\right)^2},\quad{}n - m = \text{even},
\end{equation}
\begin{equation}
Q_{mn}^\ast\left(-ic\right) = -\frac{\left(i^{-\left(m + 1\right)}k_{mn}^{\left(1\right)}\left(-ic\right)\right)^2}{c}\sum_{r = 0}^m\alpha_{r}^{mn}\left(-ic\right)\frac{\left(2m - 2r + 1\right)!}{r!\left(2^{m - r}\left(m - r\right)!\right)^2},\quad{}n - m = \text{odd},
\end{equation}
\begin{equation}
\label{x012}
\alpha_{r}^{mn}\left(-ic\right) = \left[\frac{d^r}{{dx}^r}\frac{1}{\left(\sum_{k = 0}^\infty{}c_{2k}^{mn}\left(-ic\right)x^k\right)^2}\right]_{x = 0},
\end{equation}
\begin{equation}
g_{mn}\left(-ic, i\xi\right) = \xi\left(\xi^2 + 1\right)^{-m / 2}\sum_{r = 0}^\infty{}B_{2r}^{mn}\left(-ic\right)\xi^{2r},\quad{}n - m = \text{even},
\end{equation}
\begin{equation}
g_{mn}\left(-ic, i\xi\right) = \left(\xi^2 + 1\right)^{-m / 2}\sum_{r = 0}^\infty{}B_{2r}^{mn}\left(-ic\right)\xi^{2r},\quad{}n - m = \text{odd}.
\end{equation}
To compute $\alpha_{r}^{mn}\left(-ic\right)$, we use the following procedure, which was described in \cite{zhang1996} and uses some properties of Cauchy products.
Let $C_k = c_{2k}^{mn}\left(-ic\right)$, and expand the denominator in Eq.\ (\ref{x012}) as
\begin{equation}
\left(\sum_{k = 0}^\infty{}c_{2k}^{mn}\left(-ic\right)x^k\right)^2 = \sum_{n = 0}^\infty\sum_{k = 0}^nC_kx^kC_{n - k}x^{n - k} = \sum_{n = 0}^\infty{}B_nx^n,
\end{equation}
\begin{equation}
B_n = \sum_{k = 0}^nC_kC_{n - k},
\end{equation}
\begin{equation}
\frac{1}{\displaystyle{\left(\sum_{k = 0}^\infty{}c_{2k}^{mn}\left(-ic\right)x^k\right)^2}} = \frac{1}{\displaystyle{\sum_{n = 0}^\infty{}B_nx^n}} = \sum_{n = 0}^\infty{}A_nx^n,
\end{equation}
\begin{equation}
\sum_{n = 0}^\infty{}A_nx^n\sum_{n = 0}^\infty{}B_nx^n = 1,
\end{equation}
\begin{equation}
\sum_{n = 0}^\infty\left(\sum_{k = 0}^nA_kB_{n - k}\right)x^n = 1.
\end{equation}
In order for this equality to hold,
\begin{equation}
A_0B_0 = 1,\quad\sum_{k = 0}^nA_kB_{n - k} = 0,\quad{}n > 0.
\end{equation}
Rearranging,
\begin{equation}
A_0 = \frac{1}{B_0},\quad{}A_n = -\frac{1}{B_0}\sum_{k = 0}^{n - 1}A_kB_{n - k},\quad{}n > 0.
\end{equation}
We can now compute $\alpha_{r}^{mn}\left(-ic\right)$:
\begin{equation}
\alpha_{r}^{mn}\left(-ic\right) = \left[\frac{d^r}{{dx}^r}\sum_{n = 0}^\infty{}A_nx^n\right]_{x = 0} = A_rr!.
\end{equation}
Plugging Eq.\ (\ref{x035}) into the oblate version of Eq.\ (\ref{x017}) yields the following recurrence relation in $B_{2r}^{mn}\left(-ic\right)$:
\begin{equation}
\label{x016}
\alpha_{2r}B_{2r + 2}^{mn}\left(-ic\right) + \beta_{2r}B_{2r}^{mn}\left(-ic\right) + \gamma_{2r}B_{2r - 2}^{mn}\left(-ic\right) = h_{2r},
\end{equation}
where
\begin{equation}
\alpha_{2r} = \left(2r + 2\right)\left(2r + 3\right),
\end{equation}
\begin{equation}
\beta_{2r} = \left(2r + 1\right)\left(2r - 2m + 2\right) + m\left(m - 1\right) - \lambda_{mn}\left(-ic\right),
\end{equation}
\begin{equation}
\gamma_{2r} = c^2,
\end{equation}
\begin{equation}
\label{x019}
h_{2r} = -2Q_{mn}^\ast\left(-ic\right)\left(i^{-m}k_{mn}^{\left(1\right)}\left(-ic\right)\right)^{-1}\sum_{k = r - m + 1}^\infty{}c_{2k}^{mn}\left(-ic\right)\left(m + 2k\right)\frac{\left(m + k - 1\right)!}{\left(m + k - 1 - r\right)!r!}
\end{equation}
when $n - m = \text{even}$, and
\begin{equation}
\alpha_{2r} = \left(2r + 1\right)\left(2r + 2\right),
\end{equation}
\begin{equation}
\beta_{2r} = 2r\left(2r - 2m + 1\right) + m\left(m - 1\right) - \lambda_{mn}\left(-ic\right),
\end{equation}
\begin{equation}
\gamma_{2r} = c^2,
\end{equation}
\begin{equation}
\begin{array}{cc}
\displaystyle{h_{2r} = -2Q_{mn}^\ast\left(-ic\right)\left(i^{-\left(m + 1\right)}k_{mn}^{\left(1\right)}\left(-ic\right)\right)^{-1}\left(\sum_{k = r - m}^\infty{}c_{2k}^{mn}\left(-ic\right)\left(m + 2k + 1\right)\frac{\left(m + k\right)!}{\left(m + k - r\right)!r!}\right.}\\
[-0.1in]\\
\displaystyle{\left.- \sum_{k = r - m + 1}^\infty{}c_{2k}^{mn}\left(-ic\right)\left(m + 2k\right)\frac{\left(m + k - 1\right)!}{\left(m + k - 1 - r\right)!r!}\right)}
\end{array}
\end{equation}
when $n - m = \text{odd}$.
Given a starting value, $B_0^{mn}\left(-ic\right)$, Eq.\ (\ref{x016}) can be used to compute $B_{2r}^{mn}\left(-ic\right)$.
The starting values are 
\begin{equation}
B_0^{mn}\left(-ic\right) = \left(cR_{mn}^{\left(1\right)}\left(-ic, i0\right)\right)^{-1} - Q_{mn}^\ast\left(-ic\right)R_{mn}^{\left(1\right)}\left(-ic, i0\right),\quad{}n - m = \text{even},
\end{equation}
\begin{equation}
B_0^{mn}\left(-ic\right) = -\left(cR_{mn}^{\left(1\right)\prime}\left(-ic, i0\right)\right)^{-1},\quad{}n - m = \text{odd}.
\end{equation}

\subsubsection{Calculating the Characteristic Value and Expansion Coefficients}

The procedures for computing the characteristic value and expansion coefficients for the oblate case are exactly the same as those for the prolate case, provided that $c \rightarrow -ic$.

\section{Software}

We implemented our software in C++ and MATLAB.
Our code is called \verb#spheroidal# and is primarily called via the command line.
Many programming languages have interfaces to the command line (e.g., the \verb#system# function in MATLAB), so they can access our code in this way.
There are two programs: \verb#pro_sphwv# computes the prolate spheroidal wave functions; and \verb#obl_sphwv# computes the oblate spheroidal wave functions.
Like any piece of software, our code contains many subtleties and nuances.
We have commented and documented our code to explain all of these.
Below, we describe some of the more important ones.

\subsection{Using the MPFR Library}

Except in a few places, our code uses the GNU MPFR library, which provides interfaces and routines for performing arbitrary precision arithmetic \cite{fousse2007}.
We created a C++ class, \verb#real#, which encapsulates many of the features provided by GNU MPFR.
These features include: basic arithmetic by overloading the \verb#+#, \verb#-#, \verb#*#, and \verb#/# operators; comparisons by overloading the \verb#>#, \verb#>=#, \verb#<#, \verb#<=#, \verb#==#, and \verb#!=# operators; and some elementary functions, including \verb#abs#, \verb#atan#, \verb#cos#, \verb#log#, \verb#pow#, and \verb#sin#.
GNU MPFR allows the programmer to specify the precision to use for these operations by setting the number of bits of precision.
As a reference, single and double precision arithmetic found in most programming languages have 24 and 53 bits of precision, respectively.
We experimented with several different levels of precision (as low as 24 and as high as 5000 bits of precision).
In general, as we increased the precision, the accuracy of the computations increased as well.
For lower $c$, $m$, and $n$, single and double precision were good enough.
However, for higher values, using such low precision yielded very large errors, and only by increasing the precision were these errors reduced.

\subsection{Prolate Spheroidal Wave Functions}

The code for computing the prolate spheroidal wave functions, as well as the characteristic and other special values and expansion coefficients required for computing them, is called via the command line.
Every command uses the same general structure.
The program is called \verb#pro_sphwv#.
Seven arguments are required:
\begin{enumerate}
\item{}\verb#-max_memory# is the maximum amount of memory, in MB, the program can use before automatically terminating;
\item{}\verb#-prec# is the number of bits of precision to use;
\item{}\verb#-verbose# is whether the program should output diagnostic and other information about the computations (accepted values are ``y'' and ``n'');
\item{}\verb#-c# is equal to $ka$, where $k$ is the wavenumber and $2a$ is the interfocal distance;
\item{}\verb#-m# is one modal value;
\item{}\verb#-n# is the other modal value and is equal to $m, m + 1, \ldots$; and
\item{}\verb#-w# is what the program should do.
\end{enumerate}
Following the \verb#-w# argument are zero or more arguments, the number and type of which depend on the value given for the \verb#-w# argument.
The following sequence of commands computes the characteristic and other special values and expansion coefficients required for computing the prolate spheroidal wave functions:
\begin{verbatim}
./pro_sphwv -max_memory 2000 -prec 100 -verbose y -c 10.0 -m 0 -n 0 \
     -w lambda
./pro_sphwv -max_memory 2000 -prec 100 -verbose y -c 10.0 -m 0 -n 0 -w dr \
     -n_dr 10 -dr_min 1.0e-200
./pro_sphwv -max_memory 2000 -prec 100 -verbose y -c 10.0 -m 0 -n 0 \
     -w dr_neg -n_dr_neg 10 -dr_neg_min 1.0e-200
./pro_sphwv -max_memory 2000 -prec 100 -verbose y -c 10.0 -m 0 -n 0 -w N
./pro_sphwv -max_memory 2000 -prec 100 -verbose y -c 10.0 -m 0 -n 0 -w F
./pro_sphwv -max_memory 2000 -prec 100 -verbose y -c 10.0 -m 0 -n 0 -w k1
./pro_sphwv -max_memory 2000 -prec 100 -verbose y -c 10.0 -m 0 -n 0 -w k2
./pro_sphwv -max_memory 2000 -prec 100 -verbose y -c 10.0 -m 0 -n 0 \
     -w c2k -n_c2k 10 -c2k_min 1.0e-200
\end{verbatim}
The command, \verb#./pro_sphwv ... -w lambda#, uses method (2) from Section \ref{z001} to compute the characteristic value.
Method (2) requires an approximate value for the characteristic value, which can be computed by using method (1).
We wrote a small MATLAB program for doing so.
Once this program has run, the preceeding sequence of commands can be called using one command:
\begin{verbatim}
./pro_sphwv -max_memory 2000 -prec 100 -verbose y -c 10.0 -m 0 -n 0 \
     -w everything -n_dr 10 -dr_min 1.0e-200 -n_dr_neg 10 \
     -dr_neg_min 1.0e-200 -n_c2k 10 -c2k_min 1.0e-200
\end{verbatim}
The computed values are stored as ASCII in .txt files in the \verb#data# directory.
This way, they can be reused later on without having to recompute them from scratch.
The following two commands compute the prolate spheroidal wave functions over a range of values:
\begin{verbatim}
./pro_sphwv -max_memory 2000 -prec 100 -verbose n -c 10.0 -m 0 -n 0 -w S1 \
     -a -1.0 -b 1.0 -d 0.125 -arg_type eta \
     -p 20 > data/pro_00010000_000_000_S1.txt
./pro_sphwv -max_memory 2000 -prec 100 -verbose n -c 10.0 -m 0 -n 0 -w R \
     -a 1.0 -b 9.0 -d 0.125 -arg_type xi -which R1_1,R1_2,R2_1,R2_2 \
     -p 20 > data/pro_00010000_000_000_R.txt
\end{verbatim}
where the range is determined by the values passed in for \verb#-a# (the starting point), \verb#-b# (the ending point), and \verb#-d# (the spacing between the points).
Normally, the angle functions take $\eta$ as their argument, and the radial functions take $\xi$ as their argument.
The argument, \verb#-arg_type#, allows one to use $\eta = \cos\left(x\pi\right)$ for the angle functions, where the range is over $x$, not $\eta$.
In this case, \verb#-arg_type# should be set to \verb#theta/pi#.
Likewise, for the radial functions, one can use $\xi = \left(x^2 + 1\right)^{1 / 2}$, where the range is over $x$, not $\xi$.
In this case, \verb#-arg_type# should be set to \verb#x#.
The argument, \verb#-which#, tells the program which of the different methods for computing the radial functions should be used.
The argument, \verb#-p#, is how many digits of precision to output.
There is no guarantee that all of these digits will be accurate.
How many digits are actually accurate depends on the level of precision and how many expansion coefficients are computed and used.

\subsection{Oblate Spheroidal Wave Functions}

The calling convention for \verb#obl_sphwv# is the same as \verb#pro_sphwv#.
There are some differences since a couple extra sets of expansion coefficients are computed.
The following sequence of commands computes the characteristic and other special values and expansion coefficients required for computing the oblate spheroidal wave functions:
\begin{verbatim}
./obl_sphwv -max_memory 2000 -prec 100 -verbose y -c 10.0 -m 0 -n 0 \
     -w lambda
./obl_sphwv -max_memory 2000 -prec 100 -verbose y -c 10.0 -m 0 -n 0 -w dr \
     -n_dr 10 -dr_min 1.0e-200
./obl_sphwv -max_memory 2000 -prec 100 -verbose y -c 10.0 -m 0 -n 0 \
     -w dr_neg -n_dr_neg 10 -dr_neg_min 1.0e-200
./obl_sphwv -max_memory 2000 -prec 100 -verbose y -c 10.0 -m 0 -n 0 -w N
./obl_sphwv -max_memory 2000 -prec 100 -verbose y -c 10.0 -m 0 -n 0 -w F
./obl_sphwv -max_memory 2000 -prec 100 -verbose y -c 10.0 -m 0 -n 0 -w k1
./obl_sphwv -max_memory 2000 -prec 100 -verbose y -c 10.0 -m 0 -n 0 -w k2
./obl_sphwv -max_memory 2000 -prec 100 -verbose y -c 10.0 -m 0 -n 0 \
     -w c2k -n_c2k 10 -c2k_min 1.0e-200
./obl_sphwv -max_memory 2000 -prec 100 -verbose y -c 10.0 -m 0 -n 0 -w Q
./obl_sphwv -max_memory 2000 -prec 100 -verbose y -c 10.0 -m 0 -n 0 \
     -w B2r -n_B2r 10 -B2r_min 1.0e-200
\end{verbatim}
Like in the prolate case, the command, \verb#./pro_sphwv ... -w lambda#, must be preceeded by running a small MATLAB program to compute an approximate value for the characteristic value.
The preceeding sequence of commands can be called using one command:
\begin{verbatim}
./obl_sphwv -max_memory 2000 -prec 100 -verbose y -c 10.0 -m 0 -n 0 \
     -w everything -n_dr 10 -dr_min 1.0e-200 -n_dr_neg 10 \
     -dr_neg_min 1.0e-200 -n_c2k 10 -c2k_min 1.0e-200 -n_B2r 10 \
     -B2r_min 1.0e-200
\end{verbatim}
The computed values are stored as ASCII in .txt files in the \verb#data# directory.
This way, they can be reused later on without having to recompute them from scratch.
The following two commands compute the oblate spheroidal wave functions over a range of values:
\begin{verbatim}
./obl_sphwv -max_memory 2000 -prec 100 -verbose n -c 10.0 -m 0 -n 0 -w S1 \
     -a -1.0 -b 1.0 -d 0.125 -arg_type eta \
     -p 20 > data/obl_00010000_000_000_S1.txt
./obl_sphwv -max_memory 2000 -prec 100 -verbose n -c 10.0 -m 0 -n 0 -w R \
     -a 0.0 -b 8.0 -d 0.125 -arg_type xi \
     -which R1_1,R1_2,R2_1,R2_2,R2_31,R2_32 \
     -p 20 > data/obl_00010000_000_000_R.txt
\end{verbatim}
where the arguments are the same as in the prolate case.
The only difference is that, for the radial functions, there is only one valid value for \verb#-arg_type#, which is \verb#xi#.

\subsection{Using the Wronskian}

\begin{figure}[t]
	\centering
	\includegraphics[height=2.4in]{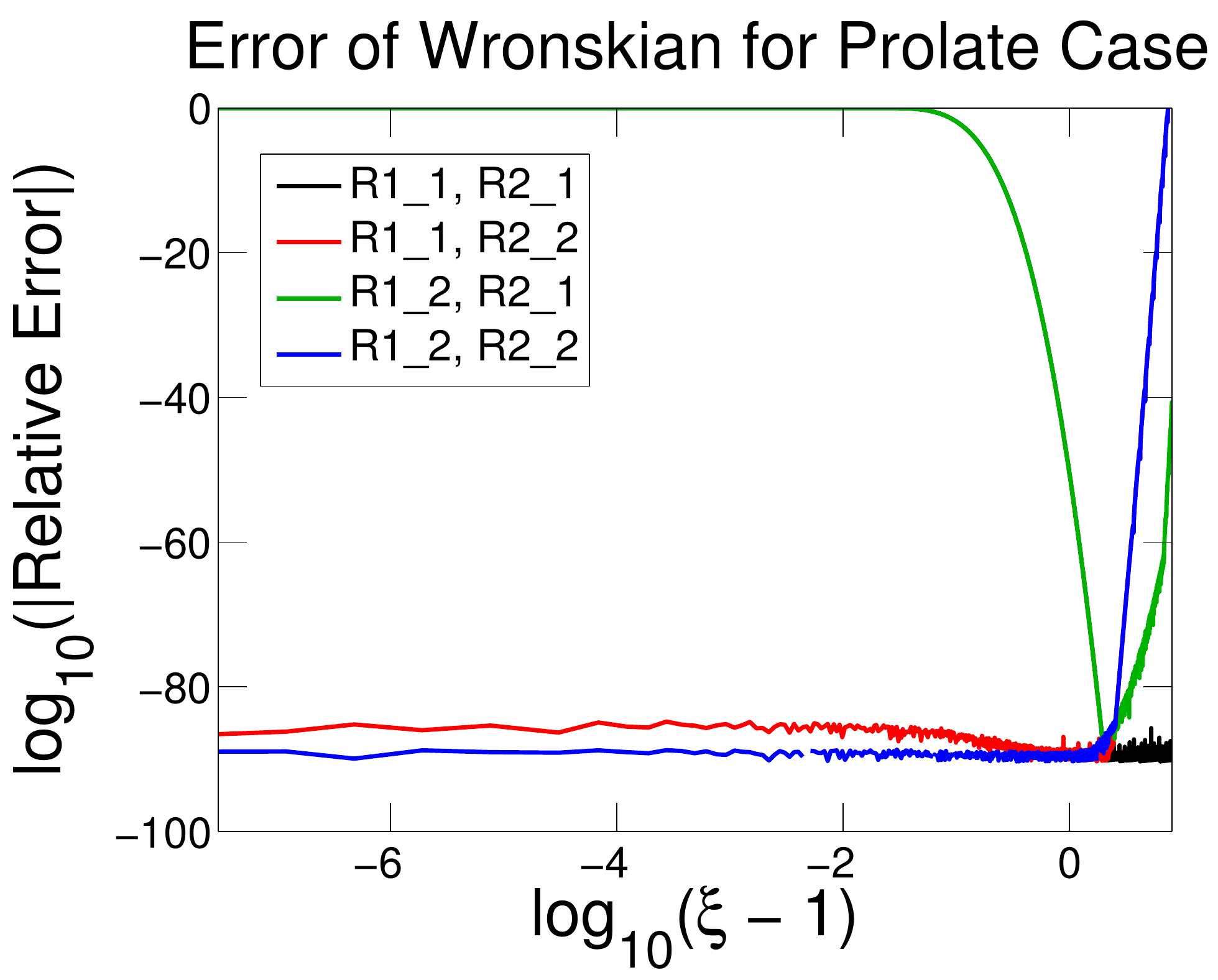}
	\caption{The relative error of the computed Wronskian when using different combinations of the methods for computing the prolate spheroidal radial functions for $c = 10$, $m = 10$, and $n = 39$.}
	\label{wronskian_compare_pro}
\end{figure}

\begin{figure}[t]
	\centering
	\includegraphics[height=2.4in]{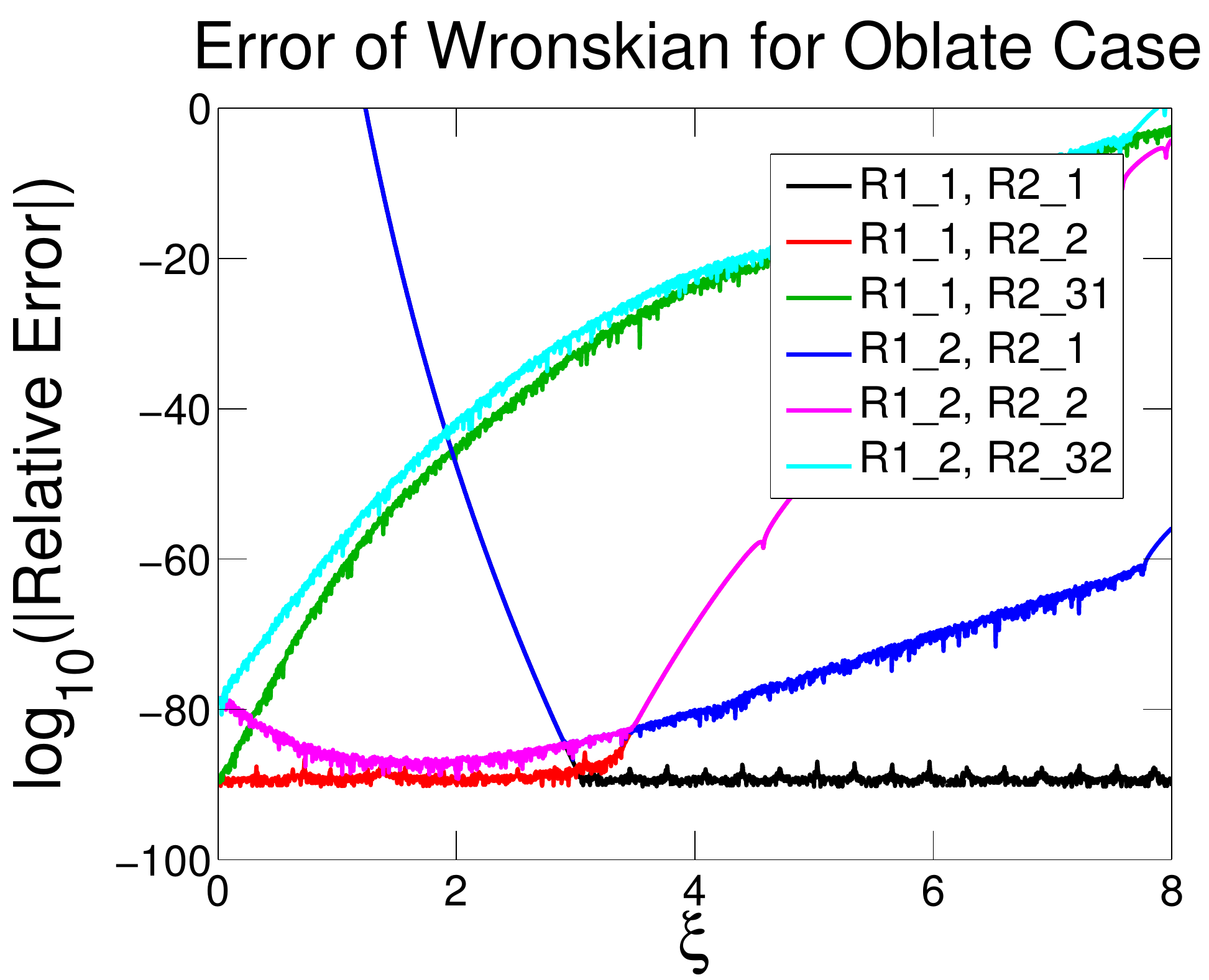}
	\caption{The relative error of the computed Wronskian when using different combinations of the methods for computing the oblate spheroidal radial functions for $c = 10$, $m = 10$, and $n = 39$.}
	\label{wronskian_compare_obl}
\end{figure}

Two methods were given for computing the prolate spheroidal radial functions of the first kind.
One method uses the spherical Bessel functions, and the other method uses a power series.
Call these methods \verb#R1_1# and \verb#R1_2#, respectively.
Likewise, two methods were given for computing the prolate spheroidal radial functions of the second kind.
One method uses the spherical Neumann functions, and the other method uses the associated Legendre polynomials of the second kind.
Call these methods \verb#R2_1# and \verb#R2_2#, respectively.
In each of these pairs of methods, one is better for certain values of $\xi$ than the other, and vice versa.
In particular, \verb#R1_1# is better for larger $\xi$ and \verb#R1_2# is better for smaller $\xi$.
Similarly, \verb#R2_1# is better for larger $\xi$ and \verb#R2_2# is better for smaller $\xi$.
However, when to use which method is not always clear: the exact value of $\xi$ below which one is better and above which the other is better is different for different values of $c$, $m$, and $n$.
We solved this dilemma in the following manner.
For a given $\xi$, all four methods are used to compute their respective functions, and the combination that yields the smallest error in the computed Wronskian is used.
An example of this can be seen in Figure \ref{wronskian_compare_pro}.
As expected, the combination, \verb#R1_2, R2_2#, was superior for smaller $\xi$ and the combination, \verb#R1_1, R2_1#, was superior for larger $\xi$.

This procedure is also used for the oblate spheroidal radial functions.
There are two methods for computing the oblate spheroidal radial functions of the first kind, one that uses the spherical Bessel functions and one that uses a power series.
Call these methods \verb#R1_1# and \verb#R1_2#, respectively.
There are three methods for computing the oblate spheroidal radial functions of the second kind, one that uses the spherical Neumann functions, one that uses the associated Legendre polynomials of the second kind, and one that uses a power series.
Call these methods \verb#R2_1#, \verb#R2_2#, and \verb#R2_3#, respectively.
Internally, \verb#R2_3# uses the oblate spheroidal radial functions of the first kind, so when \verb#R1_1# is used, call this method \verb#R2_31#, and when \verb#R1_2# is used, call this method \verb#R2_32#.
Thus, there are eight possible combinations, but only six are considered: \verb#R1_1# and \verb#R2_1#; \verb#R1_1# and \verb#R2_2#; \verb#R1_1# and \verb#R2_31#; \verb#R1_2# and \verb#R2_1#; \verb#R1_2# and \verb#R2_2#; and \verb#R1_2# and \verb#R2_32#.
For a given $\xi$, of these six, the one that yields the smallest error in the computed Wronskian is used.
An example of this can be seen in Figure \ref{wronskian_compare_obl}.
As expected, the combination, \verb#R1_1, R2_2#, was superior for smaller $\xi$ and the combination, \verb#R1_1, R2_1#, was superior for larger $\xi$.

\subsection{Solving Forward and Backward Recurrences}

\begin{figure}[t]
	\centering
	\includegraphics[height=2.4in]{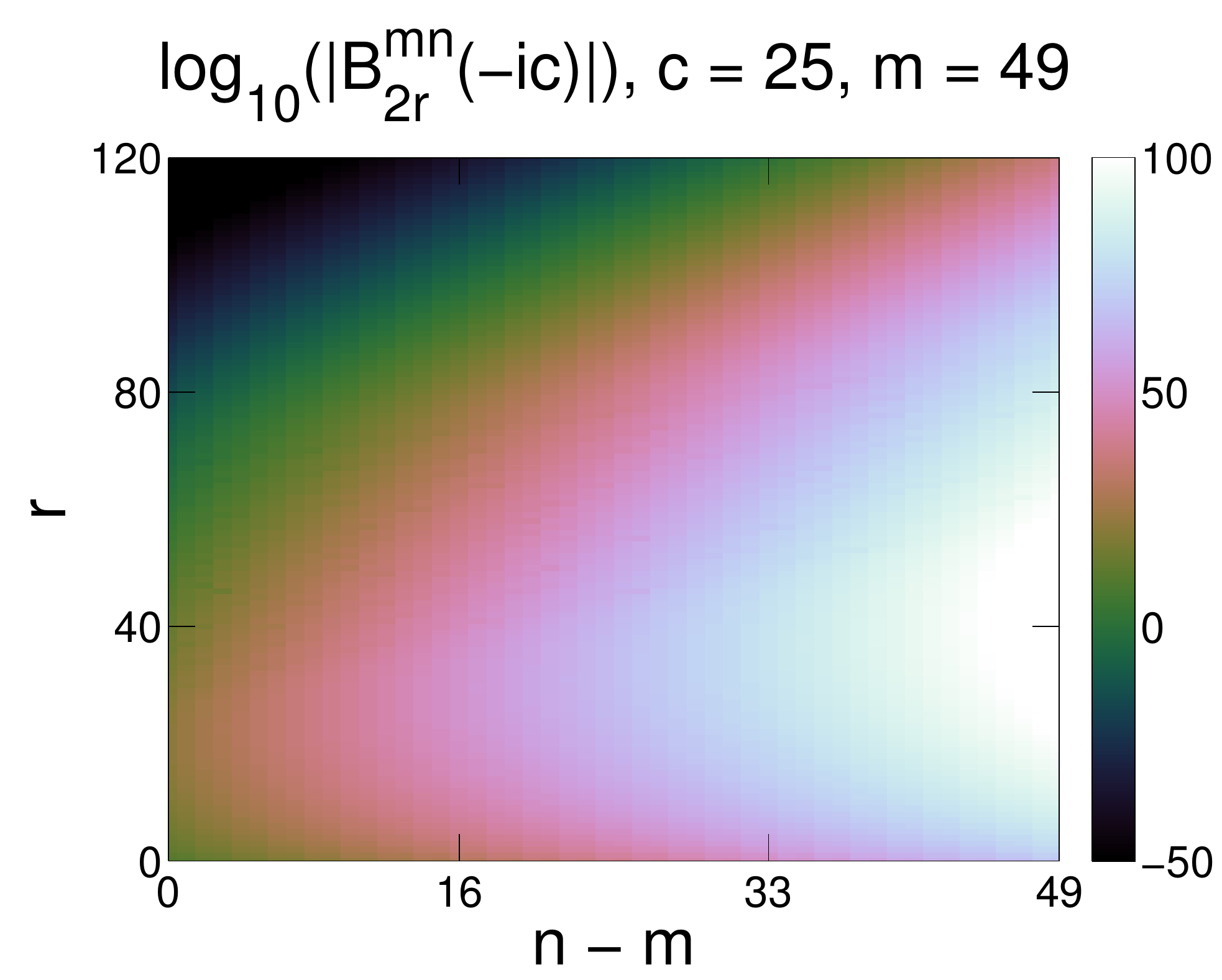}
	\caption{The order of magnitude of the expansion coefficients, $B_{2r}^{mn}\left(-ic\right)$, for $c = 25$, $m = 49$, and $n = 49, 50, \ldots, 98$.}
	\label{obl_B2r}
\end{figure}

Several recurrence relations need to be solved in order to compute the characterstic and other special values, expansion coefficients, and special functions required by many of the expressions for the spheroidal wave functions.
These recurrence relations can either have a starting value (e.g., for $B_{2r}^{mn}\left(-ic\right)$) or some kind of normalization scheme (e.g., for $d_r^{mn}\left(ic\right)$).
Except for one case, all of the recurrence relations encountered in the previous sections were homogeneous.
Depending on whether the solution to a particular recurrence relation grows or decays, the method of solution is different.
For solutions that grow (e.g., the spherical Neumann functions), the forward recurrence approach is used (i.e., the recurrence relation is used directly to compute succeeding values).
For solutions that decay (e.g., $d_r^{mn}\left(ic\right)$), the forward recurrence approach is numerically unstable.
Instead, the continued fraction approach, which is described in \cite{flammer2005}, is used.

Nonhomogeneous recurrence relations are more complicated.
For solutions that grow, the foward recurrence approach can still be used.
However, for solutions that decay, the continued fraction approach no longer works.
Instead, we use the tridiagonal matrix method described in \cite{olver1967, olver1972}: the recurrence relation is written for each index, these are combined into a tridiagonal system of equations, and this system is inverted.
The case of computing the expansion coefficients, $B_{2r}^{mn}\left(-ic\right)$, is further complicated by the fact that, for many values of $c$, $m$, and $n$, $B_{2r}^{mn}\left(-ic\right)$ grows for lower $r$, but decays for higher $r$.
See, for example, Figure \ref{obl_B2r}.
Thus, the forward recurrence approach is used when $B_{2r}^{mn}\left(-ic\right)$ is growing, and the tridiagonal matrix method is used when $B_{2r}^{mn}\left(-ic\right)$ is decaying.

\subsection{Optimizing Factorials and Rising Pochhammer Symbols}

Many of the expressions for the characteristic and other special values, expansion coefficients, and spheroidal wave functions involve infinite series expansions.
The terms in these series often include factorials and rising Pochhammer symbols.
Computing these can be rather expensive: for example, computing $a!$ and $\left(a\right)_k$ require $a - 1$ and $k - 1$ multiplies, respectively.
Because $a$ and $k$ are usually functions of the term number, as the number of terms computed in these series increases, the number of multiplies required increases quadratically.
This can be very slow.
However, this problem can be solved in the following manner: the products arising from these factorials and rising Pochhammer symbols can be factored across terms, so that only a constant number of multiplies are required for each term.

For example, consider here, again, Eq.\ (\ref{x019}):
\begin{equation}
h_{2r} = -2Q_{mn}^\ast\left(-ic\right)\left(i^{-m}k_{mn}^{\left(1\right)}\left(-ic\right)\right)^{-1}\sum_{k = r - m + 1}^\infty{}c_{2k}^{mn}\left(-ic\right)\left(m + 2k\right)\frac{\left(m + k - 1\right)!}{\left(m + k - 1 - r\right)!r!},
\end{equation}
or,
\begin{equation}
\label{x038}
h_{2r} = -2Q_{mn}^\ast\left(-ic\right)\left(i^{-m}k_{mn}^{\left(1\right)}\left(-ic\right)\right)^{-1}\sum_{k = r - m + 1}^\infty{}c_{2k}^{mn}\left(-ic\right)\left(m + 2k\right)a_k,
\end{equation}
where
\begin{equation}
a_k = \frac{\left(m + k - 1\right)!}{\left(m + k - 1 - r\right)!r!}.
\end{equation}
Computing $a_k$ requires $O\left(m + k + r\right)$ multiplies.
For large $m$, $r$, and $k$, carrying out these operations for every $a_k$ would be time consuming.
Fortunately, $a_k$ is related to $a_{k - 1}$ by a constant number of operations.
Starting with $a_{r - m + 1}$, every $a_k$ can be computed recursively and plugged into Eq.\ (\ref{x038}).
To begin,
\begin{equation}
a_{r - m + 1} = 1.
\end{equation}
For increasing $k$,
\begin{equation}
a_k = a_{k - 1}\frac{\left(m + k - 1\right)}{\left(m + k - r - 1\right)}.
\end{equation}

\section{Conclusion}

The spheroidal wave functions are among the most complicated special functions.
There are no simple ways to compute them.
However, because the solutions to so many interesting problems require them, software for computing them accurately is needed.
We have developed computational software for doing so using C++, MATLAB, and GNU MPFR, a library for performing arbitrary precision arithmetic.
In this paper, we have described the prolate and oblate spheroidal coordinate systems and wave functions, methods for deriving analytical expressions for computing them, and our software that implements these expressions.
Our software includes many novel features.
Some of these features include: using arbitrary precision arithmetic; adaptively choosing the number of expansion coefficients to compute and use; and using the Wronskian to choose from several different methods for computing the spheroidal radial functions to improve their accuracy.
We have made our software freely available on our webpage.

\bibliographystyle{jasanum}
\bibliography{spheroidal}

\begin{thebibliography}{10}
\newcommand{\enquote}[1]{``#1''}
\expandafter\ifx\csname url\endcsname\relax
  \def\url#1{\texttt{#1}}\fi
\expandafter\ifx\csname urlprefix\endcsname\relax\def\urlprefix{URL }\fi
\providecommand{\bibinfo}[2]{#2}
\providecommand{\noopsort}[1]{}
\providecommand{\switchargs}[2]{#2#1}

\bibitem{bowman1987}
\bibinfo{author}{J.~J. Bowman}, \bibinfo{author}{T.~B.~A. Senior}, and
  \bibinfo{author}{P.~L.~E. Uslenghi}, \emph{\bibinfo{title}{Electromagnetic
  and acoustic scattering by simple shapes}} (\bibinfo{publisher}{Hemisphere
  Publishing Corporation}, \bibinfo{address}{New York}) (\bibinfo{year}{1987}).

\bibitem{slepian1983}
\bibinfo{author}{D.~Slepian}, \enquote{\bibinfo{title}{Some comments on
  {F}ourier analysis, uncertainty and modeling}}, \bibinfo{journal}{SIAM
  Review} \textbf{\bibinfo{volume}{25(3)}} (\bibinfo{year}{1983}).

\bibitem{fousse2007}
\bibinfo{author}{L.~Fousse}, \bibinfo{author}{G.~Hanrot},
  \bibinfo{author}{V.~Lefevre}, \bibinfo{author}{P.~Pelissier}, and
  \bibinfo{author}{P.~Zimmermann}, \enquote{\bibinfo{title}{{MPFR}: A
  multiple-precision binary floating-point library with correct rounding}},
  \bibinfo{journal}{ACM Trans. Math. Softw.} \textbf{\bibinfo{volume}{33(2)}}
  (\bibinfo{year}{2007}).

\bibitem{flammer2005}
\bibinfo{author}{C.~Flammer}, \emph{\bibinfo{title}{Spheroidal wave functions}}
  (\bibinfo{publisher}{Dover Publications}, \bibinfo{address}{Mineola})
  (\bibinfo{year}{2005}).

\bibitem{do-nhat1996}
\bibinfo{author}{T.~Do-Nhat} and \bibinfo{author}{R.~H. MacPhie},
  \enquote{\bibinfo{title}{Accurate values of prolate spheroidal radial
  functions of the second kind}}, \bibinfo{journal}{Can. J. Phys.}
  \textbf{\bibinfo{volume}{75}} (\bibinfo{year}{1997}).

\bibitem{zhang1996}
\bibinfo{author}{S.~Zhang} and \bibinfo{author}{J.~Jin},
  \emph{\bibinfo{title}{Computation of special functions}}
  (\bibinfo{publisher}{Wiley-Interscience}) (\bibinfo{year}{1996}).

\bibitem{thompson1997}
\bibinfo{author}{W.~J. Thompson}, \emph{\bibinfo{title}{Atlas for computing
  mathematical functions}} (\bibinfo{publisher}{John Wiley and Sons})
  (\bibinfo{year}{1997}).

\bibitem{thompson1999}
\bibinfo{author}{W.~J. Thompson}, \enquote{\bibinfo{title}{Spheroidal wave
  functions}}, \bibinfo{journal}{Computing in Science and Engineering}
  \textbf{\bibinfo{volume}{1(3)}} (\bibinfo{year}{1999}).

\bibitem{vanburen2002}
\bibinfo{author}{A.~L.~V. Buren} and \bibinfo{author}{J.~E. Boisvert},
  \enquote{\bibinfo{title}{Accurate calculation of prolate spheroidal radial
  functions of the first kind and their first derivatives}},
  \bibinfo{journal}{Quart. Appl. Math.} \textbf{\bibinfo{volume}{60}}
  (\bibinfo{year}{2002}).

\bibitem{vanburen2004}
\bibinfo{author}{A.~L.~V. Buren} and \bibinfo{author}{J.~E. Boisvert},
  \enquote{\bibinfo{title}{Improved calculation of prolate spheroidal radial
  functions of the second kind and their first derivatives}},
  \bibinfo{journal}{Quart. Appl. Math.} \textbf{\bibinfo{volume}{62}}
  (\bibinfo{year}{2004}).

\bibitem{li1998}
\bibinfo{author}{L.~Li}, \bibinfo{author}{M.~Leong}, \bibinfo{author}{T.~Yeo},
  \bibinfo{author}{P.~Kooi}, , and \bibinfo{author}{K.~Tan},
  \enquote{\bibinfo{title}{Computations of spheroidal harmonics with complex
  arguments: a review with an algorithm}}, \bibinfo{journal}{Phys. Rev.}
  \textbf{\bibinfo{volume}{58}} (\bibinfo{year}{1998}).

\bibitem{falloon2003}
\bibinfo{author}{P.~E. Falloon}, \bibinfo{author}{P.~C. Abbott}, and
  \bibinfo{author}{J.~B. Wang}, \enquote{\bibinfo{title}{Theory and computation
  of the spheroidal wave functions}}, \bibinfo{journal}{J. Physics A}
  \textbf{\bibinfo{volume}{36}} (\bibinfo{year}{2003}).

\bibitem{caldwell1988}
\bibinfo{author}{J.~Caldwell}, \enquote{\bibinfo{title}{Computation of
  eigenvalues of spheroidal harmonics using relaxation}}, \bibinfo{journal}{J.
  Phys. A} \textbf{\bibinfo{volume}{21}} (\bibinfo{year}{1988}).

\bibitem{ogburn2014}
\bibinfo{author}{D.~X. Ogburn}, \bibinfo{author}{C.~L. Waters},
  \bibinfo{author}{M.~D. Sciffer}, \bibinfo{author}{J.~A. Hogan}, and
  \bibinfo{author}{P.~C. Abbott}, \enquote{\bibinfo{title}{A finite difference
  construction of the spheroidal wave functions}}, \bibinfo{journal}{Computer
  Physics Communications} \textbf{\bibinfo{volume}{185}}
  (\bibinfo{year}{2014}).

\bibitem{hodge1970}
\bibinfo{author}{D.~B. Hodge}, \enquote{\bibinfo{title}{Eigenvalues and
  eigenfunctions of the spheroidal wave equation}}, \bibinfo{journal}{J. Math.
  Phys.} \textbf{\bibinfo{volume}{11}} (\bibinfo{year}{1970}).

\bibitem{olver1967}
\bibinfo{author}{F.~W.~J. Olver}, \enquote{\bibinfo{title}{Numerical solution
  of second-order linear difference equations}}, \bibinfo{journal}{Journal of
  Research of the National Bureau of Standards} \textbf{\bibinfo{volume}{71B}}
  (\bibinfo{year}{1967}).

\bibitem{olver1972}
\bibinfo{author}{F.~W.~J. Olver} and \bibinfo{author}{D.~J. Sookne},
  \enquote{\bibinfo{title}{Note on backward recurrence algorithms}},
  \bibinfo{journal}{Mathematics of Computation}
  \textbf{\bibinfo{volume}{26(120)}} (\bibinfo{year}{1972}).

\end{thebibliography}

\end{document}